\nonstopmode

\newcommand{\ie}{i.e.{}}
\newcommand{\eg}{e.g.{}}
\newcommand{\eV}{\U{eV}}
\newcommand{\U}[1]{\,{\rm{#1}}}
\newcommand{\X}[1]{_{\mathrm{#1}}}
\newcommand{\euler}{\mathrm e}
\newcommand{\Sum}{\sum\limits}
\newcommand{\Int}{\int\limits}
\newcommand{\Lim}{\lim\limits}
\newcommand{\Max}{\max\limits}
\newcommand{\Min}{\min\limits}
\newcommand{\differential}{\>\mathrm d}
\newcommand{\E}[1]{\times 10^{#1}}
\newcommand{\XUV}{\textsc{xuv}}
\newcommand{\nbh}{\hbox{-}}
\newcommand{\xray}{x\nbh{}ray}
\newcommand{\Xray}{X\nbh{}ray}
\newcommand{\mat}[1]{\hbox{\boldmath{$#1$}\unboldmath}}
\newcommand{\mats}[1]{\hbox{\boldmath{$\scriptstyle{#1}$}\unboldmath}}
\renewcommand{\vec}[1]{\hbox{\boldmath{$#1$}\unboldmath}}
\newcommand{\eref}[1]{(\ref{#1})}
\newcommand{\Eref}[1]{Equation~(\ref{#1})}
\newcommand{\sref}[1]{section~\ref{#1}}
\newcommand{\fref}[1]{figure~\ref{#1}}
\newcommand{\Fref}[1]{Figure~\ref{#1}}
\newcommand{\tref}[1]{table~\ref{#1}}
\newcommand{\atopa}[2]{\genfrac{}{}{0pt}{}{#1}{#2}}

\documentclass[a4paper,pra,aps,10pt,twocolumn,superscriptaddress,longbibliography,showkeys,preprintnumbers]{revtex4-1}
\usepackage{CJK,amsmath,amsthm,amssymb,bbm,enumerate,color,graphicx}
\usepackage[nativepdf,bookmarks,bookmarksopen,bookmarksnumbered,raiselinks,breaklinks,%
pdftitle={Neon in ultrashort and intense x rays from free electron lasers},%
pdfauthor={Christian Buth, Randolf Beerwerth, Razib Obaid, Nora Berrah, Lorenz S. Cederbaum, Stephan Fritzsche},%
pdfsubject={Atomic Physics},%
pdfkeywords={multiphoton ionization process, x rays, photoabsorption,
Auger decay, light-matter interaction, neon atom, ion yield, radiative decay,
fluorescence, photon yield, formal language, graph, rate equations,
multiconfiguration Dirac-Hartree-Fock, cross section}]{hyperref}
\usepackage[anythingbreaks]{breakurl}
\begin{document}
\title{Neon in ultrashort and intense x rays from free electron lasers}
\author{Christian Buth}
\thanks{World Wide Web: \href{http://www.christianbuth.name}
{www.christianbuth.name}, electronic mail}
\email{christian.buth@web.de}
\affiliation{Theoretische Chemie, Physikalisch-Chemisches Institut,
Ruprecht-Karls-Universit\"at Heidelberg, Im Neuenheimer Feld~229,
69120~Heidelberg, Germany}
\author{Randolf Beerwerth}
\affiliation{Helmholtz-Institut Jena, Fr\"obelstieg~3, 07743~Jena, Germany}
\affiliation{Theoretisch-Physikalisches Institut,
Friedrich-Schiller-Universit\"at Jena, Max-Wien-Platz~1, 07743~Jena, Germany}
\author{Razib Obaid}
\affiliation{Department of Physics, University of Connecticut,
2152~Hillside Road, U-3046, Storrs, Connecticut~06269, USA}
\author{Nora~Berrah}
\affiliation{Department of Physics, University of Connecticut,
2152~Hillside Road, U-3046, Storrs, Connecticut~06269, USA}
\author{Lorenz S. Cederbaum}
\affiliation{Theoretische Chemie, Physikalisch-Chemisches Institut,
Ruprecht-Karls-Universit\"at Heidelberg, Im Neuenheimer Feld~229,
69120~Heidelberg, Germany}
\author{Stephan Fritzsche}
\affiliation{Helmholtz-Institut Jena, Fr\"obelstieg~3, 07743~Jena, Germany}
\affiliation{Theoretisch-Physikalisches Institut,
Friedrich-Schiller-Universit\"at Jena, Max-Wien-Platz~1, 07743~Jena, Germany}
\date{01 September 2017}

\begin{abstract}
We theoretically examine neon atoms in ultrashort and intense x~rays
from free electron lasers and compare our results with data from experiments
conducted at the Linac Coherent Light Source~(LCLS).
For this purpose, we treat in detail the electronic structure
in all possible nonrelativistic cationic configurations using
a relativistic multiconfiguration approach.
The interaction with the x~rays is described in rate-equation approximation.
To understand the mechanisms of the interaction, a path analysis is
devised which allows us to investigate what sequences of photoionization
and decay processes lead to a specific configuration and with what probability.
Thereby, we uncover a connection to the mathematics of graph theory and
formal languages.
In detail, we study the ion yields and find that plain rate equations
do not provide a satisfactory description.
We need to extend the rate equations for neon to incorporate
double Auger decay of a $K$\nbh{}shell vacancy and photoionization
shake off for neutral neon.
Shake off is included for valence and core ionization;
the former has hitherto been overlooked but has
important consequences for the ion yields from an \xray~energy below
the core ionization threshold.
Furthermore, we predict the photon yields from \XUV{} and \xray~fluorescence;
these allow one insights into the configurations populated
by the interaction with the x~rays.
Finally, we discover that inaccuracies in those Auger decay widths employed
in previous studies have only a minor influence on ion and photon yields.
\end{abstract}


\preprint{arXiv:1705.07521}
\maketitle

\definecolor{mablack}{rgb}{0,0,0}
\definecolor{mared}{rgb}{1,0,0}
\definecolor{malightred}{rgb}{1,0.7,0.7}
\definecolor{madarkgreen}{rgb}{0,0.5,0}
\definecolor{magreen}{rgb}{0,1,0}
\definecolor{malightgreen}{rgb}{0.7,0.85,0.7}
\definecolor{magray}{rgb}{0.4,0.4,0.4}
\definecolor{malighterred}{rgb}{1.0,0.33,0.33}
\definecolor{malightergreen}{rgb}{0.33,1.0,0.33}

\section{Introduction}

Atoms are the basic constituents of aggregates of matter realized in
molecules, clusters, and solids.
In this way, the study of the interaction of intense and ultrafast
x~rays with atoms is of fundamental importance for all research
involving matter in such light.
Intense x~rays offer manifold novel perspectives for science
such as diffraction experiments with single molecules~\cite{Neutze:BI-00}
and \xray~quantum optics~\cite{Adams:QO-13}.

Experimentally, research with intense and ultrafast x~rays has
only recently become a reality by the novel \xray~free electron lasers~(FELs)
of which there are presently four producing soft to hard x~rays:
the Linac Coherent Light Source~(LCLS)~\cite{LCLS:CDR-02,Emma:FL-10}
in Menlo Park, California, USA,
the SPring-8 Angstrom Compact free electron LAser~(SACLA)~\cite{Ishikawa:CX-12}
in Sayo-cho, Sayo-gun, Hyogo, Japan,
the SwissFEL~\cite{SwissFEL:CDR-12} in Villigen, Switzerland,
and the European X-Ray Free-Electron Laser~(XFEL)~\cite{Altarelli:TDR-06}
in Hamburg, Germany.

The novel FEL~facilities which produce x~rays with unprecedented characteristics
inspire one to investigate processes, known from the strong-field interaction of
optical lasers with atoms, but with x~rays, that go beyond the well-established
one-\xray-photon science only possible at synchrotron light
sources~\cite{Als-Nielsen:EM-01}.
With x~rays the sequential absorption of multiple photons~\cite{Hoener:FA-10,%
Fang:DC-10,Cryan:AE-10,Young:FE-10,Buth:UA-12,Fang:MI-12,Liu:RE-16,Buth:LT-18,%
Obaid:FL-18} or the simultaneous absorption of two
photons~\cite{Doumy:NA-11,Hofbrucker:RC-16} are the possible processes.
Even Rabi flopping~\cite{Rohringer:RA-08,*Rohringer:PN-08,Liu:AE-10,%
Kanter:MA-11,Demekhin:IE-11,Nikolopoulos:AC-11,Cavaletto:RF-12,Muller:RA-15}
on \xray~transitions becomes feasible.
The FEL can be used to pump
\xray~lasing~\cite{Rohringer:AI-09,*Rohringer:ER-10,Rohringer:AI-12}
on an inner-shell atomic transition which can be controlled by
an additional optical laser~\cite{Darvasi:OC-14}.
But the two-color physics of FEL x~rays and an optical laser
offers even more promising avenues: high-order harmonic generation
becomes feasible in the kiloelectronvolt regime~\cite{Buth:NL-11,Kohler:EC-12,%
Buth:KE-13,Buth:HO-15} and high-energy frequency
combs~\cite{Cavaletto:FC-13,Cavaletto:HF-14} can be produced.

Neon has been studied extensively in LCLS radiation both
experimentally~\cite{Young:FE-10,Doumy:NA-11,Obaid:FL-18} and
theoretically~\cite{Rohringer:XR-07,Ciricosta:SN-11,Xiang:RA-12,%
Gao:DD-15,Li:CR-16,Gao:SC-16}.
The quantum dynamics of neon atoms in intense and ultrashort \xray~radiation
was initially described theoretically with a plain rate-equation
model~\cite{Rohringer:XR-07,Young:FE-10,Ciricosta:SN-11} that considered only
one-photon cross sections and radiative and Auger decay widths.
The model worked well for a photon energy of~$800 \eV$ below the $K$\nbh{}shell
ionization threshold where it seemingly correctly predicted the ion
yields~\cite{Young:FE-10}.
This view did not change in a careful reinvestigation of neon in
this case using master equations~\cite{Li:CR-16}.
Yet the model failed to adequately describe ion yields for photon energies
above the $K$\nbh{}shell ionization threshold.
Subsequent theoretical research~\cite{Doumy:NA-11,Xiang:RA-12,Gao:DD-15}
revealed that there are substantial further many-electron effects in
neon in this case.
Specifically, single and double shake off following $K$\nbh{}shell
ionization of neutral neon~\cite{Doumy:NA-11} and double Auger decay
of a $K$\nbh{}shell vacancy~\cite{Gao:DD-15} were included in
extended rate-equation models for neon.
This mitigated substantially the discrepancy found between the
ion yields from the experiment and the plain rate equations.
But this was not the only shortcoming of the initial theoretical
description~\cite{Rohringer:XR-07}.
Already in Ref.~\onlinecite{Doumy:NA-11}, a remaining discrepancy between
the experimental and the theoretical ion yields was conjectured to to
be due to resonant excitations.
Such inner-shell resonant absorption~\cite{Xiang:RA-12} was
found to be essential to understand the interaction of neon
with~$1050 \eV$~x~rays.

Yet there are still gaps in our understanding of the interaction of
even one of the simplest atoms with intense and ultrafast x~rays left.
A detailed knowledge of the electronic structure of atoms
is the basis for an understanding of experiments at
\xray~FELs.
After the systematic study of neon in all cationic charge states
of Ref.~\onlinecite{Bhalla:AR-73} in the year~1973 with the
Hartree-Fock-Slater approximation, little has been done
with modern atomic theory.
On the contrary, the description of the electronic structure in
intense x-ray-atom interaction is frequently still
in Hartree-Fock-Slater~\cite{Rohringer:XR-07,Son:HA-11,*Son:EH-11}
(with relativistic energy correction~\cite{Toyota:RE-17})
or Dirac-Hartree-Slater~\cite{Buth:UA-12,Liu:RE-16,Buth:LT-18}
approximation.

This article is structured as follows.
We use state of the art atomic electronic structure theory discussed
in~\sref{sec:AES}.
Based on it, a description of the \xray~interaction with neon is
developed in~\sref{sec:xrayint} first with the photoionization cross
sections and Auger and radiative decay widths
and, second, including two-electron emission by photoionization shake off
for neutral neon and double Auger decay of a $K$\nbh{}shell vacancy in~Ne$^+$.
Ion yields are calculated in~\sref{sec:ionyields} and compared with
experimental data.
The \XUV{} and \xray~photon yields of neon are predicted in~\sref{sec:photonyi}.
Finally, conclusions are drawn in~\sref{sec:conclusion}.
In the appendices, we give detailed information on the following aspects.
To understand the involved processes in the rate equations more deeply
than with an analysis only of the Auger electron yield used
previously~\cite{Rohringer:XR-07,Young:FE-10},
we devise a path analysis in~\sref{sec:pathana}.
Computational details are given in~\sref{sec:compdet} specifying the employed
computer program packages and how integrals for photon emission
are solved efficiently numerically.

Equations are formulated in atomic units~\cite{Hartree:WM-28,Szabo:MQC-89}.
Details of the calculations in this article are provided in the
Supplementary Data~\cite{SuppData}\nocite{Mathematica:pgm-V11}.

\section{Atomic electronic structure of multiply-ionized neon}
\label{sec:AES}
\subsection{Relativistic multiconfiguration approach}

We pursue quite accurate electronic structure calculations with
state-of-the-art atomic theory~\cite{Grant:RQ-07,Froese:MC-16}.
This avails detailed fine-structure-resolved information.
Yet, nonetheless, we make the configuration approximation for the atomic
structure to simplify the description.
This refers to Hartree-Fock-Slater theory in which the
configurations of neon [Eq.~\eref{eq:configurations} below]
are the actual electronic states~\cite{Bhalla:AR-73,Froese:MC-16}.
In other words, we probabilistically average~\footnote{%
This averaging is frequently also referred to as ``statistical average''
which is a misnomer in mathematical terms.}
over the fine-structure resolved results to obtain
configuration-averaged quantities.
These are the only ones that are used subsequently.

To understand the electronic structure of the atom and photoionization,
electronic, and radiative transitions in all cationic charge states,
the cationic configurations are generated starting from the neutral
atom by systematically removing electrons.
We frequently abbreviate these configurations by their occupation numbers, \ie,
for neon in~$1s^{\ell} \, 2s^m \, 2p^n$, we simply write~$\ell m n$.
As we consider only cationic configurations, the occupation numbers
are, thereby, restricted to the maximum number found for the ground state
of the atom, \ie, the configurations of neon form the set
\begin{equation}
  \label{eq:configurations}
  \mathbb K = \{ \ell m n \mid \ell, m, n \in \mathbb N_0 \land \ell,
    m \leq 2 \land n \leq 6 \} \; .
\end{equation}

We use \textsc{Grasp2K} (General-purpose Relativistic Atomic Structure
Program)~\cite{Jonsson:2K-07,Jonsson:2K-13} to perform multiconfiguration
Dirac-Hartree-Fock~(MCDHF) computations for all~$c \in \mathbb K \land
c \neq 000$.
Typically, such a calculation is carried out for each single nonrelativistic
configuration~$\ell mn$.
However, if $\ell = m = 1$, MCDHF does not converge due to orbital
rotations~\cite{Froese:MC-16} unless the deexcited
configuration~$20n$ is included as well.
Based on the relativistic electronic structure, transitions can be determined
with \textsc{Ratip} (Relativistic Atomic Transition and Ionization
Properties)~\cite{Fritzsche:RA-12}.
Specifically, we compute photoionization, electronic, and radiative transitions.
Thereby, individual electronic structure computations
for the initial and final configurations are used.
Unfortunately, \textsc{Ratip} cannot calculate
photoionization cross sections for a hydrogen-like ion.
Thus, we switch to FAC (Flexible Atomic Code)~\cite{Gu:FA-08}
in this case.
See appendix~\sref{sec:CAESR} for details.

To make the configuration approximation for energies, widths, and
cross sections, we note that
\textsc{Grasp2K}~\cite{Jonsson:2K-07,Jonsson:2K-13} and
\textsc{Ratip}~\cite{Fritzsche:RA-12} both are fully relativistic and
fine-structure-resolved results are obtained.
That means that a nonrelativistic configuration~$c \in \mathbb K \land
c \neq 000$ does not usually correspond to a single state but to a number of
states~$N_c$, \ie, the multiplet, which are indexed in terms of the
set~$\mathbb M_c = \{1, \ldots, N_c\}$.
Therefore, a probabilistic average over the states in a multiplet is required
for the quantities of interest in configuration approximation.
To carry out this average, a uniform distribution for the states
in the multiplet is \emph{assumed}~\cite{Bhalla:TF-75}, \ie, every state
is occupied with the same probability;
thereby, we need to account for the fact that a state~$\alpha \in \mathbb M_c$
is $(2 \, J_{\alpha} + 1)$~fold degenerate with its total angular
momentum~$J_{\alpha}$.

We find the average energy of a multiplet, \ie, the energy of the
configuration~$c \in \mathbb K \land c \neq 000$, to be
\begin{equation}
  \label{eq:EC}
  E_{\mathrm{C}, c} = \dfrac{1}{S_c} \Sum_{\alpha \in \mathbb M_c}
    (2 \, J_{\alpha} + 1) \, E_{\alpha} \; ,
\end{equation}
where~$E_{\alpha}$~is the energy of the state~$\alpha \in \mathbb M_c$.
The sum of the probabilistic weights is~$S_c = \Sum_{\alpha \in \mathbb M_c}
(2 \, J_{\alpha} + 1)$.
For the electron-bare nucleus, we have~$\mathbb M_{000} = \emptyset$ and
$E_{\mathrm{C}, 000} = 0$.

To select electronic transitions and calculate transition energies and
decay widths, we need to identify potential electronic decay channels.
Thereby, lower subshell and upper subshell refer to electron subshells
which have an energy that is lower for the lower subshell and higher
for the upper subshell.
A pair of configurations is connected by an electronic decay, if
\begin{enumerate}[(a)]
  \item the final configuration has a lower subshell with an electron more than
        the initial configuration,
  \item the final configuration has two electrons less either in one
        upper subshell or distributed over two upper subshells than the initial
        configuration, and
  \item all other subshells of the initial and the final configurations are
        occupied the same.
\end{enumerate}
For each pair of configurations for which electronic transitions occur
there are initial states derived from the initial
configuration~$i \in \mathbb K$ with a higher energy than, at least one,
final state derived from the final configuration~$f \in \mathbb K$,
\ie, the maximum value of the energies of the initial-state multiplet
is larger than the minimum value of the final-state multiplet.
We sum over the difference of all such pairs of initial- and
final-state energies weighted by probabilistic factors yielding
the transition energy in configuration approximation
\begin{equation}
  \label{eq:EFI}
  E_{\mathrm{E}, f \leftarrow i} = \dfrac{1}{S'_i} \Sum_{\atopa{\scriptstyle
    \alpha \in \mathbb M_i}{\scriptstyle S''_{f\alpha} \neq 0}} (2 \,
    J_{\alpha} + 1) \, \biggl[ E_{\alpha} - \dfrac{1}{S''_{f\alpha}}
    \Sum_{\atopa{\scriptstyle \beta \in \mathbb M_f} {\scriptstyle
    E_{\beta} < E_{\alpha}}} (2 \, J_{\beta} + 1) \, E_{\beta} \biggr] \; .
\end{equation}
The sum of the probabilistic factors of the states of the final
configuration~$f$ depending on the initial state~$\alpha$
is~$S''_{f\alpha} = \Sum_{\atopa{\scriptstyle \beta \in \mathbb M_f}
{\scriptstyle E_{\beta} < E_{\alpha}}} (2 \, J_{\beta} + 1)$;
it is zero, if no final states are reachable from~$\beta$ by electronic decay.
The sum of the probabilistic factors for the initial configuration~$i$ is
given by~$S'_i = \Sum_{\atopa{\scriptstyle \alpha \in \mathbb M_i}
{\scriptstyle S''_{f\alpha} \neq 0}} (2 \, J_{\alpha} + 1)$.
At first sight, Eq.~\eref{eq:EFI} appears more complicated than it should be.
Naively, we expect that $E_{\mathrm{E}, f \leftarrow i}$ is simply given by the
difference~$E_{\mathrm{C}, f} - E_{\mathrm{C}, i}$ of the average
energies of the final and initial configurations~\eref{eq:EC}.
In fact, \Eref{eq:EFI} reduces to this case for non-overlapping multiplets
of the initial and final configuration,
\ie, if~$\Max_{\beta \in \mathbb M_f} E_{\beta} < \Min_{\alpha \in \mathbb M_i}
E_{\alpha}$ holds.
However, for overlapping multiplets, not all initial states undergo transitions
to all final states;
thus we need to pick out those transitions which actually occur and
take only those into account in the probabilistic average~\eref{eq:EFI}.
To determine the electronic decay width~$\Gamma_{\mathrm{E}, f \leftarrow i}$
in configuration approximation, we probabilistically average over the
electronic decay widths of the states
\begin{equation}
  \label{eq:GamEI}
  \Gamma_{\mathrm{E}, f \leftarrow i} = \dfrac{1}{S_i}
    \Sum_{\alpha \in \mathbb M_i}
    (2 \, J_{\alpha} + 1) \, \Sum_{\atopa{\scriptstyle \beta \in \mathbb M_f}
    {\scriptstyle E_{\alpha} > E_{\beta}}} \gamma_{\mathrm{E},
    \beta \leftarrow \alpha} \; .
\end{equation}
Here $\gamma_{\mathrm{E}, \beta \leftarrow \alpha}$~is the decay width from
the state~$\alpha$ to $\beta$.

\begin{figure*}
  \hfill(a)\includegraphics[clip,height=5cm]{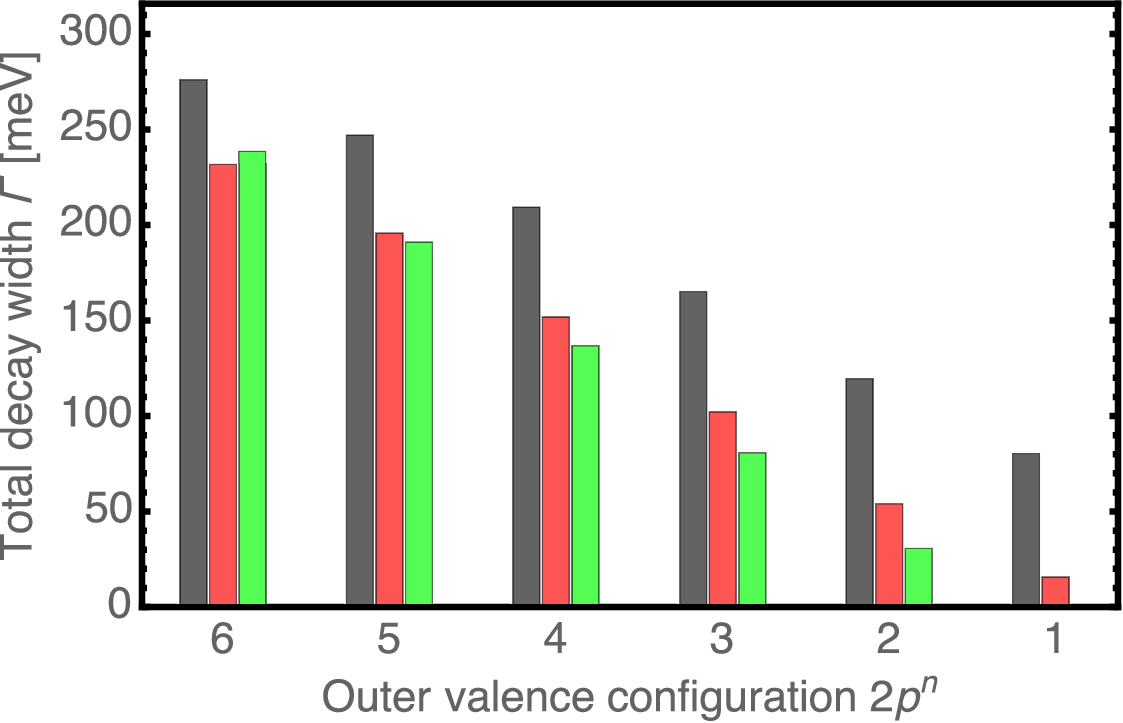}\hfill\quad\hfill
  (b)\includegraphics[clip,height=5cm]{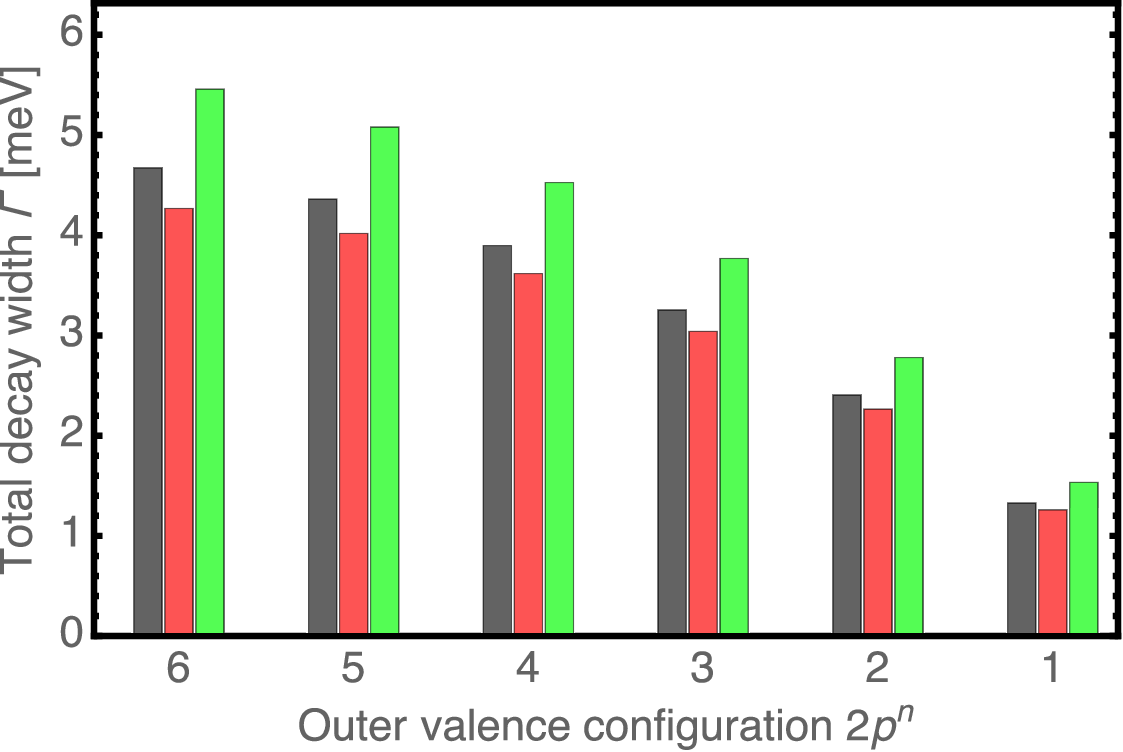}\hfill\ \ \ {}
  \caption{(Color online) Total decay widths of single core holes in neon:
            (a)~Auger widths and (b)~radiative widths.
            The \textbf{\textcolor{magray}{grey}} bars stand
            for the decay of the configurations~$1s^1 \, 2s^2 \, 2p^n$,
            \textbf{\textcolor{malighterred}{red}} bars
            for~$1s^1 \, 2s^1 \, 2p^n$, and
            \textbf{\textcolor{malightergreen}{green}} bars
            for~$1s^1 \, 2s^0 \, 2p^n$ with~$n \in \{1, \ldots, 6\}$.}
  \label{fig:sch_Auger_widths}
\end{figure*}

Radiative transitions occur between (some of) the states in the
multiplets associated with a pair of configurations, if
\begin{enumerate}[(a)]
  \item the final configuration has a lower subshell with an electron more than
        the initial configuration,
  \item the final configuration has an electrons less in an
        upper subshell than the initial configuration, and
  \item all other subshells of the initial and the final configurations are
        occupied the same.
\end{enumerate}
Then there are initial states with a higher energy than the final
states and radiative transitions occur.
All said for electronic transition energies applies also for radiative
transition energies~$E_{\mathrm{R}, f \leftarrow i}$ and we use~\eref{eq:EFI}.
Radiative decay widths~$\Gamma_{\mathrm{R}, f \leftarrow i}$ are obtained
analogously to~\eref{eq:GamEI}.

Photoionization also involves a pair of configurations chosen as follows
\begin{enumerate}[(a)]
  \item the final configuration has a subshell with an electron less than
        the initial configuration, and
  \item all other subshells of the initial and the final configurations are
        occupied the same.
\end{enumerate}
Photoionization may occur if there are initial states with a lower energy
than the final states.
The minimum photon energy required to ionize the atom in the initial
configuration is the threshold ionization energy;
it is given by
\begin{equation}
  \label{eq:EphI}
  E_{\mathrm{P}, f \leftarrow i} = \min_{\atopa{\scriptstyle \alpha \in
    \mathbb M_i, \beta \in \mathbb M_f}{\scriptstyle E_{\beta} > E_{\alpha}}}
    (E_{\beta} - E_{\alpha}) \; .
\end{equation}
This expression works also for overlapping initial and final state multiplets
because we require that $E_{\beta} > E_{\alpha}$ in the minimization
which picks out only pairs of initial and final states that can be
connected by photoionization.
The probabilistic average of the cross section
for~$f \in \mathbb K \land f \neq 000$ is
\begin{equation}
  \label{eq:xsc}
  \sigma_{f \leftarrow i} = \dfrac{1}{S_i} \Sum_{\alpha \in \mathbb M_i}
    (2 \, J_{\alpha} + 1) \Sum_{\atopa{\scriptstyle \beta \in \mathbb M_f}
    {\scriptstyle E_{\beta} > E_{\alpha}}}
    \varsigma_{\beta \leftarrow \alpha} \; ,
\end{equation}
where $\varsigma_{\beta \leftarrow \alpha}$~is the photoionization
cross section for a transition from~$\alpha$ to~$\beta$.
An empty sum in~\eref{eq:xsc} is zero.
For $f = 000$, we have~$\sigma_{000 \leftarrow i} = \tfrac{1}{S_i}
\Sum_{\alpha \in \mathbb M_i} (2 \, J_{\alpha} + 1) \, \varsigma_{0
\leftarrow \alpha}$ where~$\varsigma_{0 \leftarrow \alpha}$~stands for
the cross section to ionize the last electron of the atom in state~$\alpha$.
Here and throughout we suppress the explicit dependence of the
cross sections on the photon energy in the notation.

\subsection{Multiply-charged neon atoms}
\label{sec:multineon}

A neutral neon atom has 10~electrons;
a total of 63~nonrelativistic cationic configurations
among which 68~Auger transitions and 100~radiative transitions occur;
there are 138~one-photon ionization cross sections~\cite{SuppData}.
In~\fref{fig:sch_Auger_widths}, we depict the total Auger and radiative
decay widths of single core holes in neon which are obtained
for an~$i \in \mathbb K$ from~\eref{eq:GamEI} by summing over~$f$
for all~$f \in \mathbb K$.
For both, we observe a decreasing width with decreasing number of
$2p$~electrons as there are fewer and fewer electrons available
to make a transition.
The largest Auger width, in~\fref{fig:sch_Auger_widths}(a), is found
for~$126$, \ie, if all decay channels are present.
However, the largest radiative width, in~\fref{fig:sch_Auger_widths}(b),
occurs for the decay of~$106$, \ie, if both $2s$~electrons are missing
due to a spatially more compact trication with a resulting larger
dipole transition matrix elements involving the vacancy in the $1s$~shell
and the $2p$~subshell.
Similar trends arise for double core holes~\cite{SuppData}.

To assess the accuracy of our results for neon, we compare them to the
early calculations of Ref.~\onlinecite{Bhalla:AR-73} where the
Hartree-Fock-Slater~(HFS) approximation was used to represent the atomic
electronic structure (for details see Ref.~\onlinecite{SuppData}).
The relative error of decay widths is expressed as
\begin{equation}
  \label{eq:relerror}
  \dfrac{\Delta \Gamma}{\Gamma\X{rel}} = \dfrac{\Gamma\X{rel}
    - \Gamma\X{HFS}}{\Gamma\X{rel}} \; ,
\end{equation}
where $\Gamma\X{rel}$~stands for widths from
\textsc{Ratip}~\cite{Fritzsche:RA-12} [Eq.~\eref{eq:GamEI}]
and $\Gamma\X{HFS}$~are from Ref.~\onlinecite{Bhalla:AR-73}.
The relative error of total radiative decay widths is found to
be generally smaller for double core holes, compared with single core holes.
The reason is electron correlations in the $K$~shell for
single core holes which are not captured by the Hartree-Fock-Slater
approximation.
We find the same trend for the relative errors of the total Auger decay widths.
Also we observe that partial and total Auger decay widths from~\textsc{Ratip}
[Eq.~\eref{eq:GamEI}] are always larger than those from
Ref.~\onlinecite{Bhalla:AR-73}.
The relative error in the transition energies is small
for radiative transitions where it is in most cases smaller for double core
holes compared with single core holes.
The reverse trend is found for Auger transitions: energies for double core
hole decay are overall less accurate than for single core hole decay.
Frequently, Auger transition energies have a larger relative error than
radiative transitions energies.

\section{\Xray~interaction with neon atoms}
\label{sec:xrayint}

\begin{figure}
  \includegraphics[clip,width=\hsize]{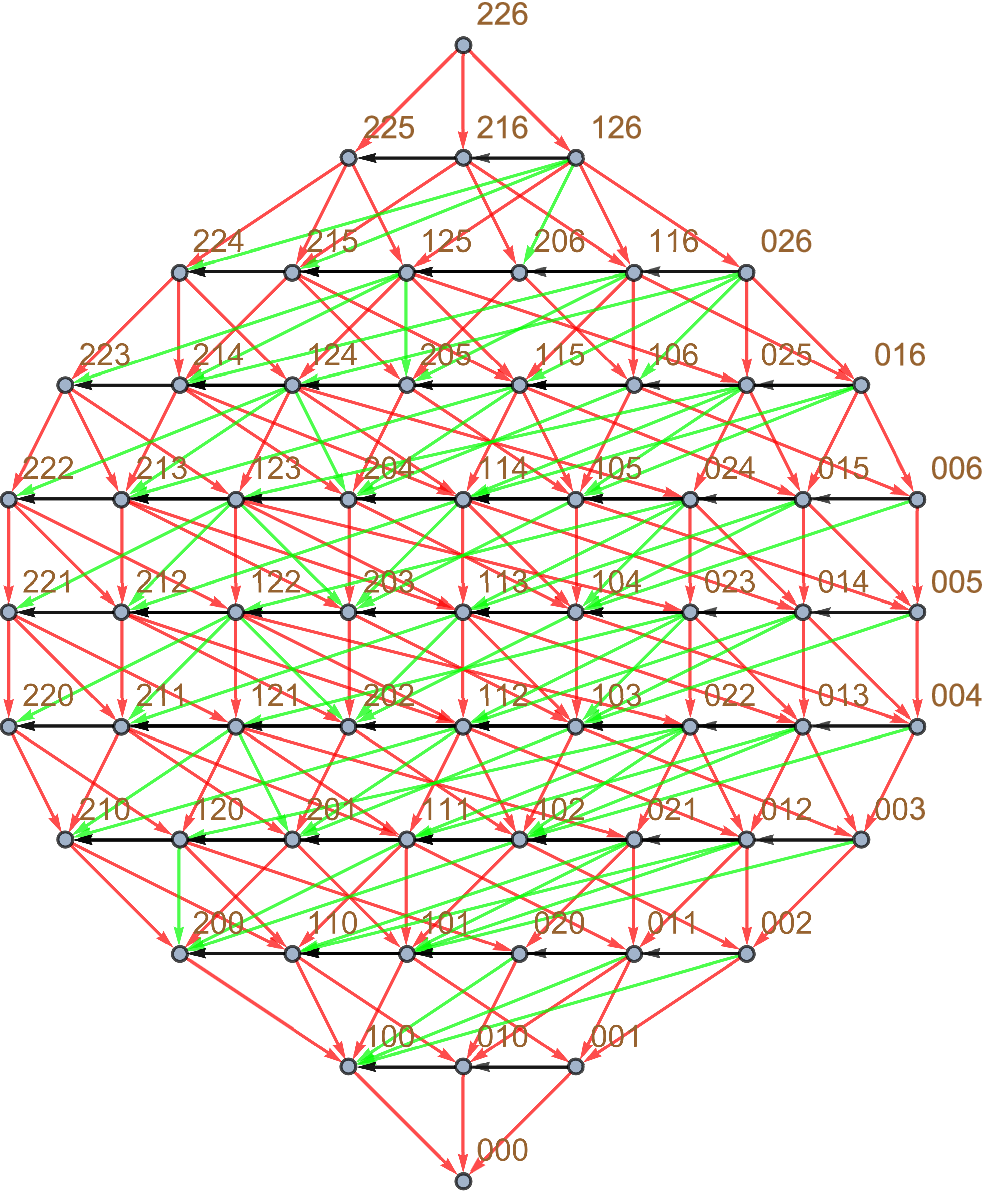}
  \caption{(Color online) Graph of the configurations of neon and transitions
           among them.
           Configurations~$1s^{\ell} \, 2s^m \, 2p^n$ are given by their
           occupation numbers~$\ell m n$.
           \textbf{\textcolor{mared}{Red}} arrows depict photoionization,
           \textbf{\textcolor{magreen}{green}} arrows Auger decay, and
           \textbf{\textcolor{mablack}{black}} arrows radiative transitions.}
  \label{fig:req_graph}
\end{figure}

The cationic configurations of neon and photoionization, Auger, and radiative
transitions among them, as debated in~\sref{sec:AES}, can be visualized
in a graph~\cite{Kozyrev:GR-16}, \fref{fig:req_graph}, similarly to
Ref.~\onlinecite{Ziaja:KB-16};
yet this gives only a static, time-independent view on the
interaction with x~rays.
To gain insights into the time-dependent quantum dynamics of the absorption
of x~rays and the resulting decay processes, we employ the rate-equation
approximation~\cite{Rohringer:XR-07,Son:HA-11,*Son:EH-11,Buth:UA-12,%
Liu:RE-16,Buth:LT-18}
that has been used successfully in a number of studies to describe
experiments, \eg, Refs.~\onlinecite{Hoener:FA-10,Young:FE-10,Doumy:NA-11,%
Ciricosta:SN-11,Xiang:RA-12,Buth:UA-12,Liu:RE-16,Gao:DD-15,Li:CR-16,%
Obaid:FL-18}.
Exemplary rate equations for a nitrogen atom restricted to $K$\nbh{}shell electrons
are given in Ref.~\onlinecite{Buth:UA-12} and the formal
structure of such rate equations is discussed in Ref.~\onlinecite{Buth:LT-18}.
Using rate equations implies that coherences are not included in the
description.
This is typically a good approximation for nonresonant absorption
of x~rays~\cite{Li:CR-16}.
However, if the \xray~energy is tuned to a resonance, then coherences
manifest~\cite{Rohringer:RA-08,*Rohringer:PN-08,Liu:AE-10,%
Demekhin:IE-11,Nikolopoulos:AC-11,Kanter:MA-11,Cavaletto:RF-12,%
Cavaletto:FC-13,Adams:QO-13,Cavaletto:HF-14,Muller:RA-15}.

\subsection{Plain rate equations}
\label{sec:plainreq}

We formulate a system of rate equations, to which we refer as plain rate
equations, using the configurations of neon~\eref{eq:configurations} and
the transitions among them determined in~\sref{sec:AES}.
The rate equations are expressed succinctly for~$j \in \mathbb K$ as follows
\begin{eqnarray}
  \label{eq:reqgam}
  \dfrac{\differential P_j(t)}{\differential t} &=& \Bigl[
    \Sum_{i \in \mathbb K} \sigma_{j \leftarrow i} \;
    P_i(t) - \sigma_j \, P_j(t) \Bigr] \, J\X{X}(t) \nonumber \\
  &&{} + \Sum_{i \in \mathbb K} \bigl( \Gamma_{\mathrm{E}, j \leftarrow i} \;
    + \Gamma_{\mathrm{R}, j \leftarrow i} \bigr) \, P_i(t) \\
  &&{} - (\Gamma_{\mathrm{E}, j} + \Gamma_{\mathrm{R}, j})
    \, P_j(t) \; , \nonumber
\end{eqnarray}
where the partial cross sections are~$\sigma_{j \leftarrow i}$ and their
sum over~$j$ for a fixed~$i$, the total cross section, is~$\sigma_i$.
The partial decay widths for Auger decay
are~$\Gamma_{\mathrm{E}, j \leftarrow i}$ and for radiative
decay are~$\Gamma_{\mathrm{R}, j \leftarrow i}$ where their sums over~$j$
for a fixed~$i$ are the total Auger and radiative decay
widths~$\Gamma_{\mathrm{E}, i}$ and $\Gamma_{\mathrm{R}, i}$, respectively.
Note that the cross sections and the decay widths are only nonzero,
if there are accessible final configurations.
The \xray~pulse is quantified in~\eref{eq:reqgam} by its photon
flux~$J\X{X}(t)$.
The initial condition at~$t \to \infty$ is that the neutral atom is in its
ground state.
In Eq.~\eref{eq:reqgam}, we suppressed the spatial coordinates for the
probabilities and the photon flux for clarity which we do also in what follows
(see appendix~\ref{sec:CAESR} for details).

\subsection{Two-electron emission in the rate equations}
\label{sec:manyreq}

There are substantial many-electron effects in neon that are not
captured by the cross sections and Auger decay widths considered
so far in the rate equations~\eref{eq:reqgam}.
A lot of effort has gone into investigating such many-electron effects in
neutral neon and singly ionized neon manifesting themselves as
photoionization shake off and double Auger decay, respectively.
However, information about such two-electron emission in higher-charged
neon cations is missing.
Specifically photoionization shake off is significant
for~Ne$^+$ because of its large contribution in neutral neon.
However, generally, two-electron emission becomes less
important the more-highly charged the cation is as the
ionization threshold for liberating an additional electron
moves considerably upward in energy~\cite{SuppData}.
In consequence, we extend the plain rate equations from~\sref{sec:plainreq}
using the experimental results from
Refs.~\onlinecite{Saito:SO-94,Kanngiesser:RN-00,Hikosaka:MA-15} to include
photoionization shake off and double Auger decay~\cite{SuppData}.

\subsubsection{Modified rate equations due to photoionization shake off}

For a photon energy larger than about~$1050 \eV$---well above the double
ionization threshold for a $K$\nbh{}shell electron and a valence
electron---photoionization shake off is saturated~\cite{Saito:SO-94}.
Then the number of doubly charged atoms produced by photoionization---for
which not only the $K$\nbh{}shell electron is ejected but, additionally, a
valence electron leaves the atom---is 23\%~of the number of atoms
produced with a single $K$~vacancy [table~III of
Ref.~\onlinecite{Saito:SO-94}].
For photon energies just above the neon $K$~edge, this percentage
varies, however, as, \eg, there is insufficient energy to eject a
valence electron from the $2s$~subshell or at all.
This can be disregarded in what follows because the photon energy is
always much larger than the neon $K$~edge or below it.
We need to relate this experimental ratio of double ionization
to single ionization~$r_{1sL/1s} = 0.23$ to the core-ionization
cross section~$\sigma_{126 \leftarrow 226}$
which already is in the plain rate equations~\eref{eq:reqgam}.
For this purpose, we assume that $\sigma_{126 \leftarrow 226}$
provides the total amount of $K$~vacancies formed which needs to
be corrected for shake off, \ie, the contribution of single ionization is
calculated with~$(1 - s\X{C}) \, \sigma_{126 \leftarrow 226}$ and the
contribution of double ionization follows
from~$s\X{C} \, \sigma_{126 \leftarrow 226}$
where $s\X{C}$~obeys~$\tfrac{s\X{C}}{1 - s\X{C}} = r_{1sL/1s}
\Longleftrightarrow s\X{C} = \tfrac{r_{1sL/1s}}{1 + r_{1sL/1s}}$.
Furthermore, we assume that only a $2p$~electron is ejected by shake off.
The plain rate equations are modified as follows where we abbreviate
the time derivative by~$\dot\  = \tfrac{\differential}{\differential t}$:
\begin{widetext}
\begin{eqnarray}
  \label{eq:P126}
  \dot P_{126}(t) &=& (1 - s\X{C}) \, \sigma_{126 \leftarrow 226} \, J\X{X}(t)
    \, P_{226}(t) - \sigma_{126} \, J\X{X}(t) \, P_{126}(t)
    - \Gamma_{\mathrm{E}, 126} \, P_{126}(t) - \Gamma_{\mathrm{R}, 126}
    \, P_{126}(t) \\
  \label{eq:P125}
  \dot P_{125}(t) &=& s\X{C} \, \sigma_{126 \leftarrow 226} \, J\X{X}(t) \,
    P_{226}(t) + \sigma_{125 \leftarrow 225} \, J\X{X}(t) \, P_{225}(t)
    + \sigma_{125 \leftarrow 126} \, J\X{X}(t) \, P_{126}(t)
    - \sigma_{125} \, J\X{X}(t) \, P_{125}(t) \; , \\
  &&{} - \Gamma_{\mathrm{E}, 125} \, P_{125}(t)
    + \Gamma_{\mathrm{R}, 125 \leftarrow 116} \, P_{116}(t)
    + \Gamma_{\mathrm{R}, 125 \leftarrow 026} \, P_{026}(t)
    - \Gamma_{\mathrm{R}, 125} \, P_{125}(t) \; . \nonumber
\end{eqnarray}
\end{widetext}
There is also the possibility that two valence electrons are liberated
in the course of $K$\nbh{}shell ionization for a high enough \xray~energy.
However, the ratio of triple ionization to single ionization is
only~0.02 [table~III of Ref.~\onlinecite{Saito:SO-94}] which
is too small to have a noticeable impact on the
ion yields or photon yields in this article.
Consequently, this process is neglected in what follows.

Along the lines of the preceding paragraph, photoionization shake off
in the course of valence ionization~\cite{Saito:SO-94} is treated.
Namely, multiple photoionization of valence electrons occurs
if the x~rays have sufficient energy to double-valence-ionize the atom.
For x~rays above~$250 \eV$ the ratio of double to single ionization is
saturated and amounts to~$16\%$ [table~I of Ref.~\onlinecite{Saito:SO-94}].
Likewise to~\eref{eq:P126}, we modify the rate equations for~$225$ and $216$;
analogous to~\eref{eq:P125} the rate equations for~$224$ [Eq.~\eref{eq:P224}
below] and $215$ are changed.
Triple and quadruple valence ionization with the saturated ratios of~0.014
and 0.002, respectively, are not included because they are very small.

\subsubsection{Modified rate equations due to double Auger decay}
\label{sec:DA}

Double Auger decay~\cite{Amusia:DA-92,Kabachnik:CC-07} occurs for
Ne$\,1s$~holes---where two electrons are emitted instead of one---with
a probability of~$5.97\%$ of all $K$\nbh{}shell~holes [table~I of
Ref.~\onlinecite{Kanngiesser:RN-00}];
normal Auger decay takes place for~$92.67\%$ of the decays.
Triple Auger decay makes only a negligible contribution
of~0.38\% and is thus not considered here.
The remaining probability is due to radiative decay producing~Ne$^+$;
this is treated already in our plain rate equations~\eref{eq:reqgam}.
Hence, if an electronic decay takes place, double Auger decay occurs with
a probability of~$d = \tfrac{0.0597}{0.0597+0.9267} = 0.06052$
and normal Auger decay with probability~$1 - d = 0.9395$~\footnote{%
Here $d$~includes direct and cascade double Auger decay~\cite{Amusia:DA-92};
direct decay refers to the simultaneous emission of two electrons;
cascade decay of a Ne$\,1s$~hole means that there is first normal
Auger decay of the vacancy to a short-lived excited state of~Ne$^{2+}$ for which
the $2s$~subshell is not fully occupied and one electron is in a
Rydberg orbital
(see Refs.~\onlinecite{Ueda:DF-03,Yoshida:ST-05,DeFanis:IV-05,Kitajima:IV-06});
then a second electronic decay takes place producing the final state
of~Ne$^{3+}$.
We do not distinguish these two types of double Auger decay but treat
them jointly because the relevant excited states of~Ne$^{2+}$ neccessary
to describe cascade decay are not incorporated in the plain
rate equations~\eref{eq:reqgam}.}.
From the probability of~$5.97\%$ for double Auger decay of a
$K$\nbh{}shell vacancy in~Ne$^+$, one obtains the double Auger
partial probabilities for the production of final configurations of~Ne$^{3+}$
[column ``Total~DA'' in table~I of Ref.~\onlinecite{Hikosaka:MA-15}];
they are~$p_{2p^{-3}} = 2.5/100$, $p_{2s^{-1} \, 2p^{-2}} = 2.9/100$, and
$p_{2s^{-2} \, 2p^{-1}} = 0.5/100$ for the distribution of the
three defect electrons over the valence shell of~Ne$^{3+}$.
The total is~$p\X{DA} = p_{2p^{-3}} + p_{2s^{-1} \, 2p^{-2}}
+ p_{2s^{-2} \, 2p^{-1}} = 0.059$.
Double Auger decay opens up an additional channel to normal Auger decay.
Therefore, we need to introduce an extra electronic decay term in the
plain rate equations~\eref{eq:reqgam}.
Yet the emission of the second electron in double Auger decay is triggered
by the emission of the first electron~\cite{Amusia:DA-92} and thus not
independent of the width for normal Auger decay.
We find the fractions~$d_{223 \leftarrow 126} = \tfrac{p_{2p^{-3}}}
{p\X{DA}} \, d$, $d_{214 \leftarrow 126} = \tfrac{p_{2s^{-1} \, 2p^{-2}}}
{p\X{DA}} \, d$, and $d_{205 \leftarrow 126} = \tfrac{p_{2s^{-2} \, 2p^{-1}}}
{p\X{DA}} \, d$.
They quantify the amount of~$\Gamma_{\mathrm{E}, 126}$ which leads to
the configurations~$223$, $214$, or $205$.
Furthermore, we let~$d_{126} = d_{223 \leftarrow 126} + d_{214 \leftarrow 126}
+ d_{205 \leftarrow 126}$.
To describe double Auger decay, we modify the rate equations via
\begin{widetext}
\begin{eqnarray}
  \label{eq:P224}
  \dot P_{224}(t) &=& (1 - d_{126}) \Gamma_{\mathrm{E}, 224 \leftarrow 126} \,
    P_{126}(t) + \sigma_{224} \, J\X{X}(t) \, P_{224}(t)
    + \sigma_{224 \leftarrow 225} \, J\X{X}(t) \, P_{225}(t) \\
  &&{} + s\X{V} \, \sigma_{225 \leftarrow 226} \, J\X{X}(t) \, P_{226}(t)
    + \Gamma_{\mathrm{R}, 224 \leftarrow 125} \, P_{125}(t)
    + \Gamma_{\mathrm{R}, 224 \leftarrow 215} \, P_{215}(t) \nonumber \\
  \label{eq:P223}
  \dot P_{223}(t) &=& d_{223 \leftarrow 126} \, \Gamma_{\mathrm{E}, 126}
    \, P_{126}(t) + \sigma_{223 \leftarrow 224} \, J\X{X}(t) \, P_{224}(t)
    - \sigma_{223} \, J\X{X}(t) \, P_{223}(t)
    + \Gamma_{\mathrm{E}, 223 \leftarrow 125} \, P_{125}(t) \\
  &&{} + \Gamma_{\mathrm{R}, 223 \leftarrow 214} \, P_{214}(t)
    + \Gamma_{\mathrm{R}, 223 \leftarrow 124} \, P_{124}(t) \; . \nonumber
\end{eqnarray}
\end{widetext}
Likewise to Eq.~\eref{eq:P224}, the rate equations for~$215$ and $206$
are changed, the one for~$206$, however, without valence shake off
term ``$s\X{V}$''.
Analogous to Eq.~\eref{eq:P223}, we modify the rate equations for~$214$
and $205$.

\section{Ion yields of neon atoms}
\label{sec:ionyields}

Rate equations [\sref{sec:xrayint}] provide time-dependent probabilities
to find an atom in a configuration.
Such a time-dependent population is difficult to measure.
Therefore, ion yields are considered instead which are derived from the
probability at infinite time to find the atom in a specific charge state
renormalized to the probability to find an ion then at
all~\cite{Buth:UA-12,Liu:RE-16}.
Ion yields are experimentally accessible via the ion
time-of-flight~\cite{Hoener:FA-10,Young:FE-10}.
Furthermore, they have the decisive advantage that, unlike probabilities,
they do not depend on the geometry of the interaction volume or the
gas density.
In what follows, we compare theoretical ion yields from our
rate equations [\sref{sec:xrayint}] with previously obtained
theoretical and experimental data for neon in LCLS
radiation~\cite{Young:FE-10,Doumy:NA-11}.

\begin{figure}
  \includegraphics[clip,width=\hsize]{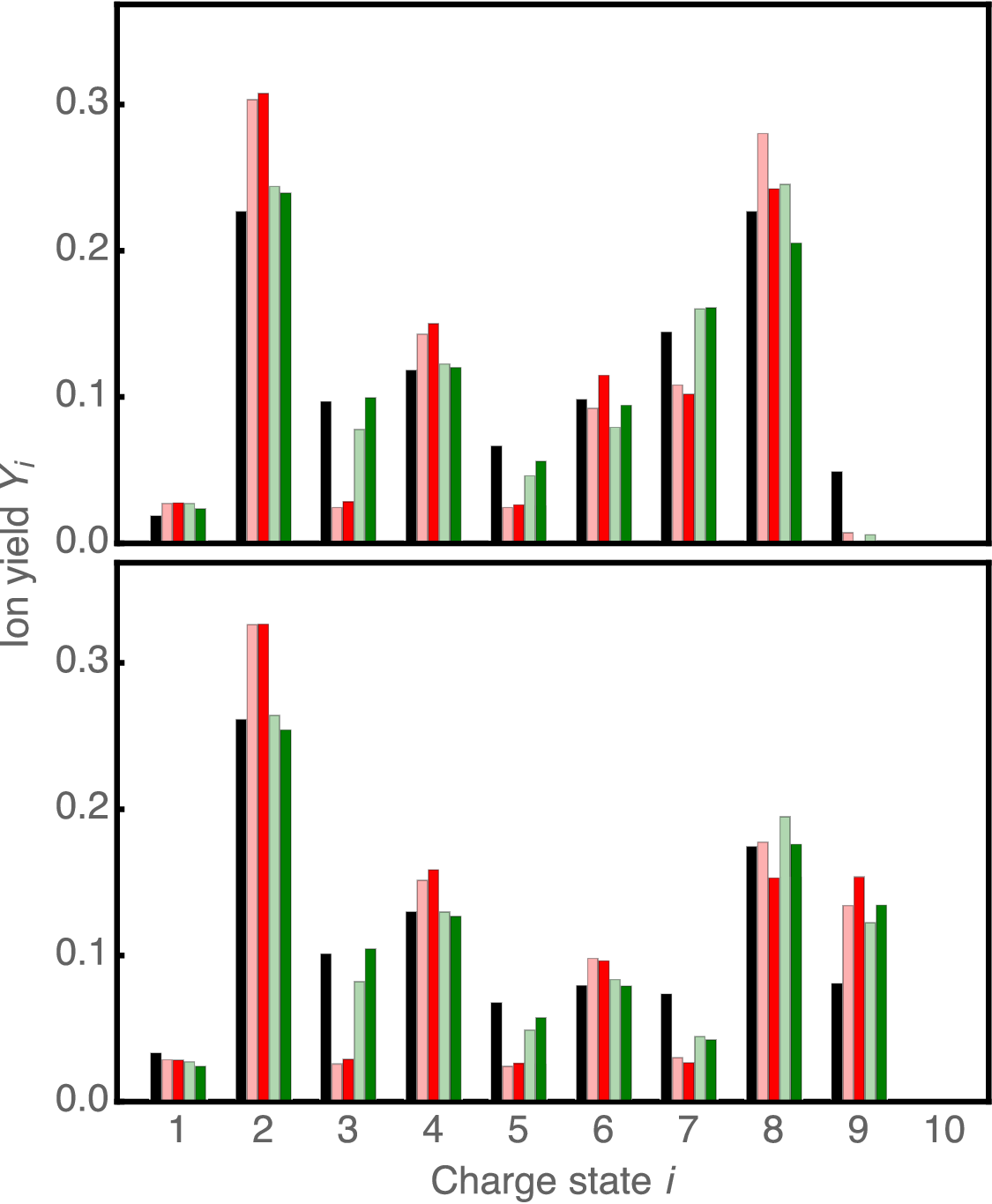}
  \caption{(Color online) Ion yields of neon atoms in LCLS~radiation at photon
           energies of~$1110 \eV$ (upper panel) and of~$1225 \eV$ (lower panel).
           \textbf{\textcolor{mablack}{Black}} bars represent experimental
           data from figure~2 in Ref.~\onlinecite{Doumy:NA-11}
           for a FWHM pulse duration of~$100 \U{fs}$.
           The nominal pulse energies are~$1.27 \U{mJ}$ (upper panel) and
           $1.45 \U{mJ}$ (lower panel).
           In both cases, we assume a reduction of the nominal pulse energy
           by a factor of~$0.699/3$ due to transmission losses
           in the \xray~optics;
           this factor is not specified in Ref.~\onlinecite{Doumy:NA-11}
           but is taken from Ref.~\onlinecite{Young:FE-10} for
           a photon energy of~$1050 \eV$.
           The Gaussian beam is taken to have a FWHM major axis of~$2 \U{\mu m}$
           and a FWHM minor axes of~$1 \U{\mu m}$ as in
           Ref.~\onlinecite{Young:FE-10};
           in Ref.~\onlinecite{Doumy:NA-11} only an approximate beam area
           of~${\sim}2 \U{\mu m^2}$ is noted.
           \textbf{\textcolor{malightred}{Light red}} and
           \textbf{\textcolor{malightgreen}{light green}} bars are the
           theoretical ion yields from figure~2 in Ref.~\onlinecite{Doumy:NA-11}
           without and with inclusion of shake off, respectively.
           \textbf{\textcolor{mared}{Red}} bars show our predictions with
           the plain rate equations [\sref{sec:plainreq}] and
           \textbf{\textcolor{madarkgreen}{green}} bars stand for the ion yields
           from the modified rate equations [\sref{sec:manyreq}].}
  \label{fig:doumy}
\end{figure}

In~\fref{fig:doumy}, we show the ion yields of neon for the
\xray~energies of~$1110 \eV$ (upper panel) and $1225 \eV$ (lower panel).
There is almost no yield for~Ne$^{10+}$ as the photon energy is, in both
cases, below the core-ionization threshold of~Ne$^{9+}$.
We compare the theoretical ion yields from
our plain rate equations [\sref{sec:plainreq}] to the equivalent in
Ref.~\onlinecite{Doumy:NA-11} which were obtained
with the rate equations from Ref.~\onlinecite{Rohringer:XR-07}
and find the results of both computations to be very similar.
There is a pronounced asymmetry between even and odd charge states
in the theoretical values apart from~Ne$^{9+}$ in the lower panel.
Namely, the ion yields for odd charge states are generally much lower
than for even charge states.
Based on the path analysis of appendix~\ref{sec:pathana},
this is because the dominant paths in the plain rate equations
are mainly composed of sequences of~``CA''~processes [\tref{tab:maxprob}]
which cause an even number of electrons to be removed from
the atom and are the principal contribution to produce even charge states.
For the odd charges states, mostly an additional valence ionization needs
to occur, \ie, the string starts with~``V'' [\tref{tab:maxprob}], which
suppresses such paths.
However, this asymmetry between odd and even charge states in the
theoretical ion yields is not as pronouncedly reflected in
the experimental ones;
apart from~Ne$^+$, for the odd charges states, we find that
the measured ion yields are drastically underestimated by the
plain rate equations apart from~Ne$^{9+}$ in the lower panel.
This poor agreement of theory with experiment has already been noted in
Ref.~\onlinecite{Young:FE-10}.
It hints at the omission of further ionization channels that
cause an increase of the odd charge states.

The ion yields from the modified rate equations [\sref{sec:manyreq}]
are also presented in~\fref{fig:doumy}.
We observe a large improvement over the ion yields from the plain rate
equations [\sref{sec:plainreq}] leading to a good agreement with the
experimental data for most charge states.
Specifically, the theoretical ion yields from our modified rate equations
are very similar to those obtained with the theory of
Ref.~\onlinecite{Doumy:NA-11} that includes theoretical calculations
of single and double shake off after $K$\nbh{}shell ionization.
This mitigates the previously found asymmetry between odd and even
charge states because single photoionization shake off for~Ne---the only one
considered by us in~\sref{sec:manyreq}---removes an additional electron over
the primary ionization.
Thereby, shake off increases the amount of~Ne$^{2+}$ and,
if a $K$\nbh{}shell hole is created, makes upon Auger decay a
contribution to~Ne$^{3+}$.
Likewise, double Auger decay of a $K$\nbh{}shell vacancy in~Ne$^+$
produces~Ne$^{3+}$.
However, the agreement between theory and experiment is not
always good, namely, for the ion yield of~Ne$^{7+}$ (lower panel)
and of~Ne$^{9+}$ (both panels).
The Ne$^{9+}$~ion yield in the upper panel is far too low because simultaneous
two-photon ionization~\cite{Doumy:NA-11} is not included in our
modified rate equations.
The discrepancy in the~Ne$^{7+}$~ion yield in the lower panel
can be ascribed to an inner-shell resonant excitation~\cite{Doumy:NA-11}
which is also missing in our description.
Such resonant processes have a significant impact on the
ion yields for an \xray~energy of~$1050 \eV$~\cite{Xiang:RA-12} (see below).

\begin{figure}
  \includegraphics[clip,width=\hsize]{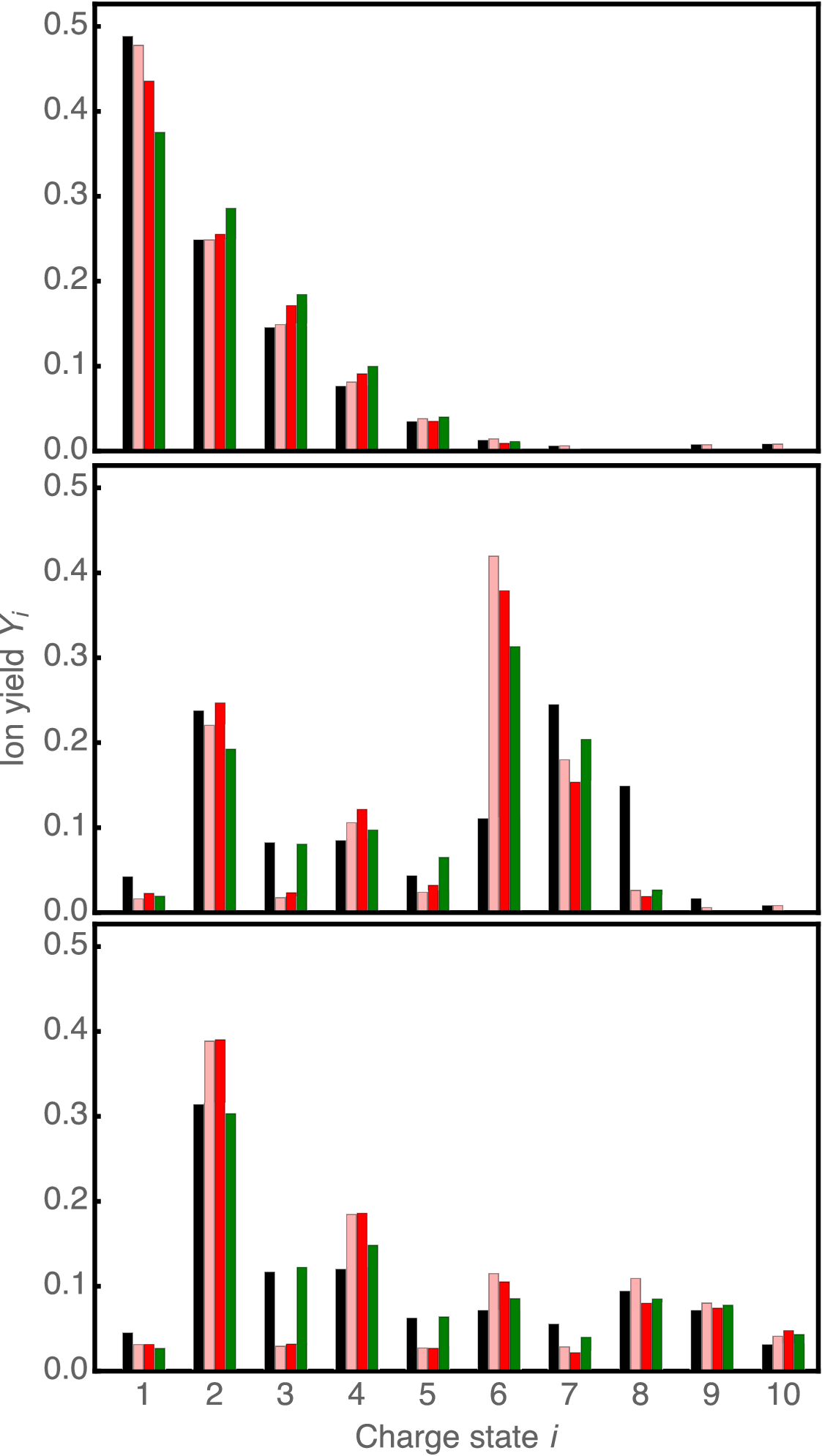}
  \caption{(Color online) Ion yields of neon atoms in LCLS~radiation at
           photon energies of~$800 \eV$ (upper panel), of~$1050 \eV$
           (middle panel), and of~$2000 \eV$ (lower panel).
           Experimental data stems from Fig.~2b
           in Ref.~\onlinecite{Young:FE-10}.
           From top panel to bottom panel,
           the FWHM pulse durations are~$340 \U{fs}$, $280 \U{fs}$, and
           $230 \U{fs}$, respectively;
           the nominal pulse energy is~$2.4 \U{mJ}$ which needs to be
           multiplied by the factors~$0.663 / 3$, $0.699 / 3$, and
           $0.584 / 3$, respectively, to obtain the actual pulse
           energies~\cite{Young:FE-10}.
           Our prediction for~$2000 \eV$~photon energy, however, is made using
           a factor of~$0.25$ instead of~$0.584 / 3 = 0.195$
           and a FWHM \xray~beam diameter of~$1 \U{\mu m}$.
           See text for further details.
           Everything else as in~\fref{fig:doumy}.}
  \label{fig:young}
\end{figure}

In~\fref{fig:young}, we display the ion yields of neon for the
\xray~energies of~$800 \eV$ (upper panel), $1050 \eV$ (middle panel),
and $2000 \eV$ (lower panel).
In the upper panel, ion yields from the theory of
Ref.~\onlinecite{Rohringer:XR-07} are compared with results
from our plain rate equations [\sref{sec:plainreq}] and very good
agreement between both and the experimental data is observed~\footnote{%
Yet puzzling is that basically no Ne$^{8+}$~ion yields were measured but
Ne$^{9+}$ and Ne$^{10+}$.
Our theory predicts no noticeable values for Ne$^{7+}$~and higher charge states;
neither do the computations of Refs.~\onlinecite{Ciricosta:SN-11,Li:CR-16}
for charge states~Ne$^{8+}$ and higher.}.
Here the \xray~energy is below the $K$\nbh{}shell ionization threshold
where photoionization shake off with respect to core ionization
and double Auger decay do not occur.
Yet there is still photoionization shake off from valence ionization,
an effect that has hitherto been disregarded.
Our modified rate equations [\sref{sec:manyreq}] showcase the importance
of this effect which lifts the good agreement between theory and experiment.
In the middle and lower panels, we have a situation very much like
the one in~\fref{fig:doumy} and large parts of its discussion is
relevant here as well.
As before, the plain rate equations give rise to ion yields which
are in stark discrepancy from the ones found in the experiment.
The inclusion of photoionization shake off and double Auger decay
in the modified rate equations leads to an overall satisfactory
agreement between theory and experiment.
However, specifically in the middle panel, there is a
drastic discrepancy discovered for~Ne$^{6+}$ and Ne$^{8+}$.
In Ref.~\onlinecite{Xiang:RA-12}, this case is investigated
and the observed deviations are ascribed to resonant excitations
which are not included in our modified rate equations.
For the lower panel, we use different parameters than those
in Ref.~\onlinecite{Young:FE-10}.
Otherwise no agreement---not even with the published theoretical
ion yields and our plain rate equations---is obtained.
Our choice of parameters produces a peak intensity
of~$2.2 \E{17} \U{\tfrac{W}{cm^2}}$ which is consistent
with~$4.5 \E{17} \U{\tfrac{W}{cm^2}}$ at a FWHM pulse
duration of~$140 \U{fs}$ used in Ref.~\onlinecite{Gao:DD-15} to
produce similar theoretical ion yields as the ones shown in the lower panel
with the modified rate equations.
To achieve about the intensity of Ref.~\onlinecite{Gao:DD-15} we reduced the
FWHM spot size of the \xray~beam.
Without reducing the spot size, we would need a factor of~$0.635$
instead of~$0.25$ to determine the actual pulse energy from the
nominal one;
we consider this to be unrealistically high.
Certainly, reducing the spot size of the \xray~beam significantly
is questionable as well because the design limit of the \xray~optics
is specified in Ref.~\onlinecite{Young:FE-10} to be just our
value of a FWHM beam diameter of~$1 \U{\mu m}$.

Overall, we note that there is a good agreement between the theory
of this work and the one of Refs.~\onlinecite{Young:FE-10,Doumy:NA-11}.
The observed variations do not originate from differences in the Auger decay
widths between ours and the ones from Ref.~\onlinecite{Bhalla:AR-73}
on which the theoretical computations of Ref.~\onlinecite{Young:FE-10} are
based;
they turn out to have a small influence~\cite{SuppData}.
This is ascribed to the dominance of ``CA''\nbh{}style paths
[\tref{tab:maxprob}] instead of double core holes in the interaction
with the x~rays because a precise timing of the Auger decay is not so
important then.
We use the nominal pulse duration in all calculations, however, the
actual pulse duration was measured to be much shorter
in experiments~\cite{Young:FE-10,Dusterer:FS-11}.
As noticed in~\cite{Buth:UA-12}, there is only a small influence
on the ion yields, if, \eg, the pulse duration is halved, again,
because of the prevalence of ``CA''\nbh{}style paths.
There still remains uncertainty about the actually available pulse energy for
the experimental data of Ref.~\onlinecite{Young:FE-10,Doumy:NA-11}.
In Ref.~\onlinecite{Ciricosta:SN-11} the authors report that the pulse
energy needed to be increased compared with the values of
Ref.~\onlinecite{Young:FE-10} to reproduce the theoretical ion yields there.
In principle, we expect that the actual pulse energy at the sample
should be obtainable by comparing the experimental Ne$^+$~ion yield computed
considering only the amount of~Ne$^+$ and Ne$^{2+}$~ionization
with the theoretical one from the modified rate equations [\sref{sec:manyreq}]
which include all relevant physical effects.
Yet this approach does not work in practice~\cite{SuppData}, seemingly,
as the experimental parameters, such as the spot size of the beam,
are insufficiently well known.
In Ref.~\onlinecite{Gao:DD-15}, the consequences of direct double Auger
decay of a $K$\nbh{}shell vacancy in~Ne$^+$ is examined for the prediction
of the ion yields in the lower panel of~\fref{fig:young};
there, it is claimed that shake off due to Auger decay was treated
in~\cite{Doumy:NA-11} which, however, is not the case
(knock off is the more important contribution~\cite{Gao:DD-15}).
The impact of double Auger decay of a Ne$\,1s$~hole is also found to be
significant by us~\cite{SuppData}---it chiefly influences the ion yields
of Ne$^{2+}$ and higher odd charge states---just as
direct double Auger decay studied in Ref.~\onlinecite{Gao:DD-15}.

\section{Photon yields of neon atoms}
\label{sec:photonyi}

\begin{figure*}
  \includegraphics[clip,height=5cm]{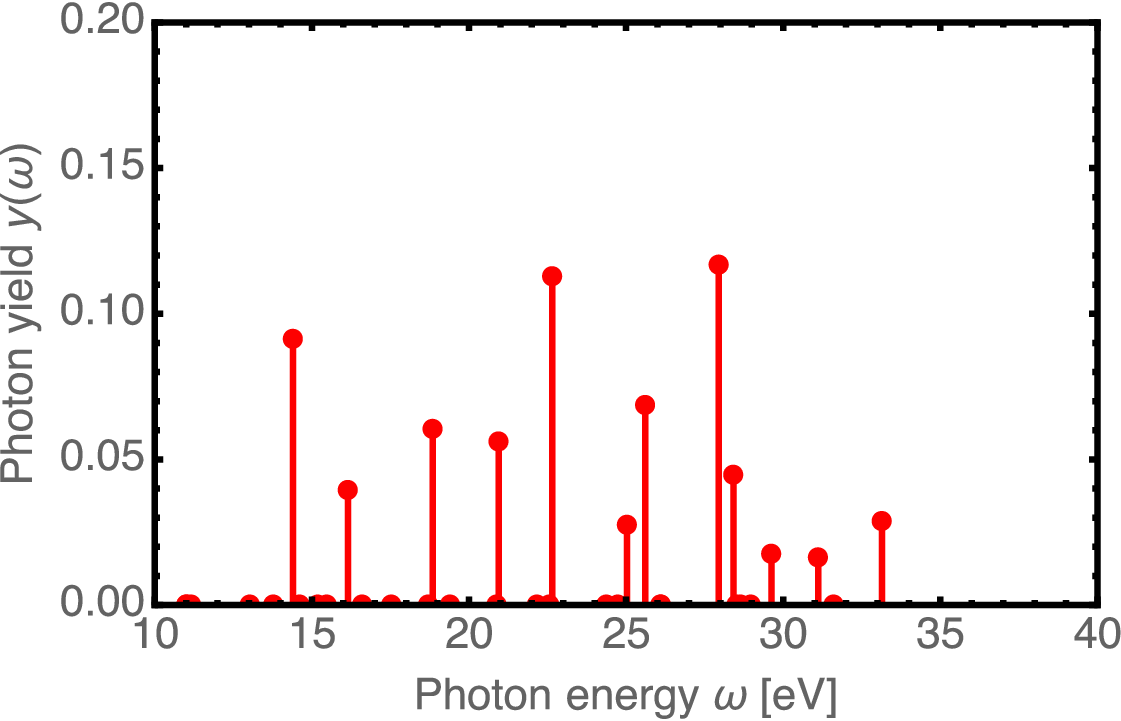}\qquad
  \includegraphics[clip,height=5cm]{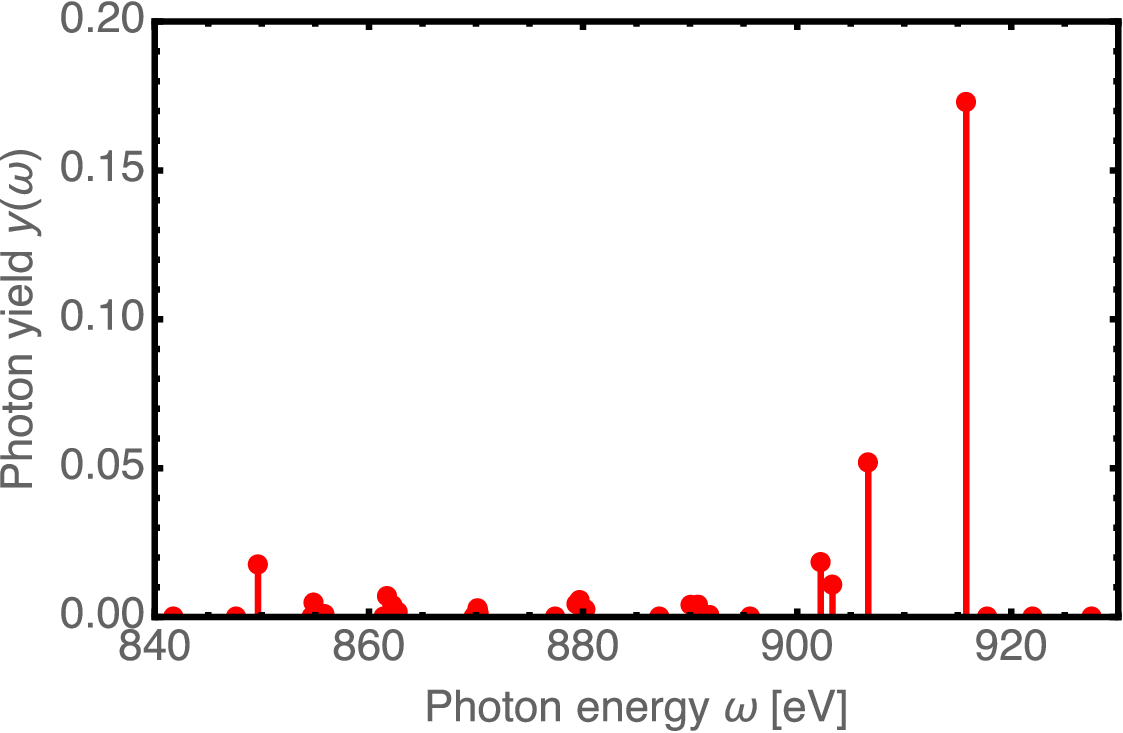}
  \caption{(Color online) Photon yield~\eref{eq:photspec} of neon in the
           \XUV{}~(left) and the \xray~(right)~regime.
           The pulse energy is~$2.4 \U{mJ}$ with $25\%$~transmission,
           the photon energy is~$1200 \eV$, and the pulse duration (FWHM)
           is~$100 \U{fs}$.
           The Gaussian beam has a FWHM major axis of~$2 \U{\mu m}$ and
           a FWHM minor axes of~$1 \U{\mu m}$.}
  \label{fig:fluorescence_yields}
\end{figure*}

Studying the fluorescence of multiply-charged atoms allows one
deep insights into the involved states in the radiative transitions.
This provides much more detailed information compared with ion yields
[\sref{sec:ionyields}].
Fluorescence spectroscopy complements Auger electron
spectroscopy~\cite{Kabachnik:CC-07,Cavaletto:RF-12};
the latter has been used so far in LCLS
experiments~\cite{Cryan:AE-10,Fang:DC-10}.
A drawback of fluorescence spectroscopy is the low fluorescence
yield for light elements [\sref{sec:multineon}].

To predict fluorescence spectra based on rate equations, we
calculate the probability for photon emission~$F_{j \leftarrow i}(t)$~up
to time~$t$ for the transition~$i \to j$ with~$i,j \in \mathbb K$
[Eq.~\eref{eq:configurations}], via
\begin{equation}
  \label{eq:photonem}
  \dfrac{\differential}{\differential t} F_{j \leftarrow i}(t)
    = \Gamma_{\mathrm{R}, j \leftarrow i} \, P_i(t) \; .
\end{equation}
Initially, we have~$\Lim_{t \to -\infty} F_{j \leftarrow i}(t) = 0$~\footnote{%
\Eref{eq:photonem} is analogous to Eq.~\eref{eq:decayprob} for~$k = \mathrm{R}$
and, in fact, $F_{j \leftarrow i}(t) = \tfrac{\Gamma_{\mathrm{R},
j \leftarrow i}}{\Gamma_{\mathrm{R}, i}} \> Q_{k, i}(t)$ holds,
if $i$~is not a terminal configuration, \ie, $\Gamma_{\mathrm{R}, i} \neq 0$.}.
Instead of~\eref{eq:photonem}, a Monte Carlo algorithm for rate equations
can be used as for xenon in Ref.~\onlinecite{Son:MC-12,*Son:EM-15}.
The probability for photon emission is experimentally only very difficult
to access because it depends on the precise geometry of the interaction
volume and the gas density within it.
Theoretically, this implies that the probability depends on the size of
the beam cross section we integrate over.
From the probability for fluorescence photon emission,
we calculate the easily experimentally accessible
photon yield~$y_{j \leftarrow i}$ for the radiative
transition~$i \rightarrow j$ which is independent of the detailed
characteristics of the interaction volume just like the ion yields
from~\sref{sec:ionyields};
it is defined by
\begin{equation}
  y_{j \leftarrow i} = \Lim_{t \to \infty} \dfrac{F_{j \leftarrow i}(t)}
    {\Sum_{k, l \in \mathbb K} F_{l \leftarrow k}(t)} \; ,
\end{equation}
for~$i, j \in \mathbb K$.
In the denominator, we sum over all radiative transitions, \ie,
we calculate the total probability for fluorescence photon emission.
The photon yield is expressed as a spectrum via
\begin{equation}
  \label{eq:photspec}
  y(\omega) = \Sum_{j, i \in \mathbb K} y_{j \leftarrow i} \>
    \delta_{\omega \, E_{\mathrm{R}, j \leftarrow i}} \; ,
\end{equation}
and the Kronecker-$\delta$~\cite{Kuptsov:KD-16} where
$E_{\mathrm{R}, j \leftarrow i}$~is the radiative transition energy
analogous to Eq.~\eref{eq:EFI}.

The photon yields of neon are displayed in~\fref{fig:fluorescence_yields}.
Several strong \XUV{} and \xray~lines are discernible.
Those radiative transition which have no competing Auger decay
exhibit the strongest photon yield, \eg, the strongest \xray~line
at~$915.80 \eV$ stems from the radiative transition~$101 \to 200$.
Particularly, all \XUV~lines originate from the exclusively radiative
decay of configurations with fully occupied $1s$~shell, \eg, the
strongest \XUV~line at~$27.95 \eV$ is from~$215 \to 224$.
As the \xray~pulse is spatially a Gaussian beam, there are large
areas of its cross section where the associated \xray~flux is comparatively
low such that, at most, one x~ray is absorbed which produces predominantly
a single core hole.
Hence the fairly strong fluorescence on the $126 \to 225$~transition
at~$849.65 \eV$ despite the competing Auger decay.
Recording the \XUV~lines in an experiment reveals rich information
about the interaction of the atom with x~rays because it indicates
the population of a specific configuration which is produced by the \xray~pulse
as radiative decay is so slow that hardly any takes place during an
ultrashort \xray~pulse.
This allows one insights into the kind of \xray~processes that have occurred.
Another advantage of also recording the \XUV~spectrum is that one has simply
more lines to measure compared with the \xray~spectrum as there are
12~radiatively-decaying configurations without competing Auger decay
in the former case compared with~4 in the latter one
[see Eq.~\eref{eq:radconfig} below].
However, \fref{fig:fluorescence_yields} reveals that in spite of competing
Auger decay, more \xray~lines, nonetheless, are relevant than
predicted by this simple argument.
We observe that inaccuracies in the Auger decay widths have only a small
effect on photon yields~\cite{SuppData}.

\section{Conclusion}
\label{sec:conclusion}

\Xray~FELs such as LCLS, SACLA, SwissFEL, or XFEL offer novel prospects for
atomic physics in intense and ultrafast x~rays.
For the understanding of experiments at FELs, a thorough knowledge
of the atomic electronic structure of the sample is essential.
In this work, we lay the foundation for its theoretical treatment
with \textsc{Grasp2K} and electronic and radiative transitions
as well as photoionization cross sections with \textsc{Ratip}
in terms of the \textsc{caesr}~program that processes all nonrelativistic
cationic configurations of an atom.
The time-dependent quantum dynamics of the interaction of x~rays with
an atom is described in rate equation approximation.
First, plain rate equations are formulated which contain only the cross sections
and decay widths from \textsc{caesr}.
Second, modified rate equations are developed that treat additionally
double Auger decay of a $K$\nbh{}shell vacancy in~Ne$^+$ and photoionization
shake off for neutral neon.
A detailed path analysis is devised to unravel the mechanisms that
produce certain atomic configurations by \xray~absoption and decay.
Specifically, the configurations and transitions among them are
arranged in terms of a graph.
The interaction with the x~rays is then characterized by paths
from the neutral atom to a cationic configuration.
The transitions in the paths are classified in terms of an alphabet
which allows us to assign a string to a path, \ie, words of a formal language.
We find that sequential few-photon absorption is the prevailing mechanism
as proposed before.
From the solution of the rate equations, we derive easily experimentally
accessible quantities, the ion yields and photon yields.
We obtain theoretical results for neon atoms in LCLS x~rays
and compare with previously obtained experimental data.
We find that the plain rate equations do not provide a satisfactory description
of the ion yields of neon, not even below the $K$\nbh{}shell ionization
threshold, a fact that has been overlooked in previous investigations.
Instead, the modified rate equations are required which lead to
an overall good agreement apart from charge states for certain photon energies
which are significantly populated also by inner-shell resonant absorption.
Photon yields from \XUV{} and \xray~fluorescence are predicted
and shown to provide more detailed information about the interacting atom
than ion yields.
We discover that inaccuracies in those Auger decay widths employed
in previous studies have only a minor influence on ion and photon yields.

Rich prospects for novel research are enabled by our study.
With our program \textsc{caesr}, other atoms can be examined
and more detailed knowledge of the interaction with x~rays
shall be gained.
The foundation of \textsc{caesr} on state-of-the-art relativistic
multiconfiguration methods for atoms provides a solid basis to describe
further processes in configuration approximation such as double Auger
decay, photoionization shake off, and resonances.
Our approach even allows one to go beyond the configuration approximation
and to consider also fine-structure-resolved atomic states and transitions.
For the ion yields of neon, there is still not always good agreement between
the theoretical prediction and the experimental data potentially due to,
yet unincorporated---photoionization shake off is also relevant in
higher charge states of neon---or unknown, many-electron effects in neon
cations.
Future investigations of such effects are very important
to ensure that our understanding of the underlying physics is
complete.
In principle, our improved theory for the interaction of neon
with x~rays allows one an accurate determination of the actual pulse energy at
the sample, a quantity which is still not very well known in LCLS experiments
due to uncertainties in the pulse diagnostics and losses in the \xray~optics.
Namely, the experimental ion yield of singly ionized neon---determined only
from singly and doubly ionized neon---should match the theoretical
value, if all other experimental parameters are accurately determined
because our theoretical understanding of the population of these
two charge states is accurate.
However, our comparisons between theory and experiment reveal that
this is not the case to date.
Photon yields offer exciting novel possibilities for future examinations
complementary to Auger electron yields.
Specifically, the combined measurement of \XUV{} and \xray~spectra
facilitate one to unravel at least some the configurations populated
by the absorption of x~rays and the ensuing Auger decay processes.

\begin{acknowledgments}
We thank Jochen Schirmer and Sang-Kil Son 
for helpful discussions and a critical reading of the manuscript.
N.B.~and R.O.~acknowledge support from the U.S.~Department of Energy,
Office of Science, Basic Energy Sciences, Division of Chemical Sciences,
Geosciences, and Biosciences under grant No.~DE-SC0012376.
L.S.C.~acknowledges support by the U.S.~Army Research Laboratory and the
U.S.~Army Research Office under Grant No.~W911NF-14-1-0383.
This research has been funded by the German Bundesministerium f{\"u}r Bildung
und Forschung under Contract No.~05K16SJA.
\end{acknowledgments}

\appendix

\section{Path analysis}
\label{sec:pathana}
\subsection{Weights and probabilities of paths}
\label{sec:probabilities}

The plain rate equations~\eref{eq:reqgam} tell us how probability flows
through the graph~\cite{Kozyrev:GR-16} in~\fref{fig:req_graph}.
However, from the solution of~\eref{eq:reqgam}, we cannot easily identify
what predominant mechanisms are at work in the interaction with the x~rays.
In Ref.~\onlinecite{Young:FE-10}, stepwise absorption of \xray~photons
by core electrons~``C''~\footnote{%
In Refs.~\onlinecite{Rohringer:XR-07,Young:FE-10}, the letter~``P''
was used to indicate single and double photoionization of core electrons
in neon.}
with subsequent Auger decay~``A'' is proposed
as prevailing mechanism, if energetically allowed.
Otherwise only valence ionization~``V'' occurs.
This inference is based on the Auger electron yield~\cite{Rohringer:XR-07}.
In what follows, we analyze in detail this proposition and, thereby,
devise an analysis to elucidate the processes responsible for
reaching a configuration and the probability for doing so.

\Fref{fig:req_graph} shows a simple and directed graph~\cite{Kozyrev:GR-16};
simple graph means that multiple edges and loops are disallowed
and directed graph indicates that the edges have orientations.
If the \xray~energy is high enough such that all configurations
in~\fref{fig:req_graph} can be reached, it is a connected graph, \ie,
it contains one directed path between every pair of vertices;
otherwise it is a disconnected graph as certain configurations
cannot be populated.
The vertices are the cationic configurations of the
atom~\eref{eq:configurations};
they are arrange from top to bottom in rows where each row corresponds
to a specific charge state of the atom.
Within a row the configurations are sorted from left to right with
increasing energy.
The edges in the graph are the transitions.
We denote configurations that are not subject to electronic or radiative decay
as terminal configurations.
The atom remains in such a configuration unless photoionization takes place.
In the graph in~\fref{fig:req_graph}, the terminal configurations are
at the very left, \ie, they are the energetically lowest configuration
in each charge state and thus there is one terminal configuration per
charge state.

Looking at~\fref{fig:req_graph}, we face a complicated situation;
typically, there are many paths that start at the neutral atom and lead
to the same last configuration.
We need to identify those ones which are important, \ie, dominant paths.
To do so, first, all possible paths are required and, second,
these paths need to be assigned probabilities to assess their
contribution to reach a specific configuration.
All possible paths are generated with a recursive
algorithm~\cite{Uspenskii:AL-11}.
The length of a path is the number of configurations it traverses.
The path of length one is the path that starts and also ends in the
configuration of the neutral atom.
The paths of length two start from the neutral atom and contain an additional
photoionization [\fref{fig:req_graph}].
The paths of length three append either another photoionization or
a decay process to the paths of length two.
This algorithm continues to generate paths of
increasing length until the electron-bare configuration of the atom is reached.
To limit the total number of paths, we only consider
radiative decay in paths, if no competing Auger decay is present.
As radiative decay is so much slower than Auger
decay, \fref{fig:sch_Auger_widths}, the omitted
paths do not carry a noticeable probability for the analysis to follow.

So far we know only all possible paths in the graph~\cite{Kozyrev:GR-16}
in~\fref{fig:req_graph} connecting the neutral atom with any other
configuration.
Yet there is a difference between the importance of
a photoionization, Auger decay, or radiative decay to occur, \ie,
a specific probability is associated with each process depending also
on the configuration from which they originate and its population during
the interaction with the x~rays.
Depending on the number and kind of transitions in a path, it
has a certain weight for reaching a specific configuration.
To find these weights, we start by obtaining the probabilities of an
atom in configuration~$i \in \mathbb K \land i \neq 000$ to undergo
photoionization or decay in~\fref{fig:req_graph} from the
probabilities~$P_i(t)$ of~\eref{eq:reqgam}.
This corresponds to taking the terms on the right-hand side of~\eref{eq:reqgam}
which cause a reduction of the rate, \ie, those with a minus sign.
To determine the probability for photoionization, we integrate
\begin{equation}
  \dfrac{\differential}{\differential t} Q_{\mathrm{P}, i}(t)
    = \sigma_i \, P_i(t) \, J\X{X}(t) \; ,
\end{equation}
and for decay processes, we integrate
\begin{equation}
  \label{eq:decayprob}
  \dfrac{\differential}{\differential t} Q_{k, i}(t)
    = \Gamma_{k, i} \, P_i(t) \; ,
\end{equation}
with~$k \in \{\mathrm{E}, \mathrm{R}\}$ where ``$\mathrm{E}$''~denotes
electronic decay and ``$\mathrm{R}$''~stands for radiative decay.
Initially, we have~$\Lim_{t \to -\infty} Q_{k, i}(t) = 0$
with~$k \in \{\mathrm{P}, \mathrm{E}, \mathrm{R}\}$.
From the probabilities~$Q_{k, i}(t)$, we determine branching ratios
\begin{equation}
  B_{k, i} = \Lim_{t \to \infty} \dfrac{Q_{k, i}(t)}{Q_{\mathrm{P}, i}(t)
    + Q_{\mathrm{E}, i}(t) + Q_{\mathrm{R}, i}(t)} \; ,
\end{equation}
which give the fraction of atoms in the configuration~$i$ that
undergo a process~$k \in \{\mathrm{P}, \mathrm{E}, \mathrm{R}\}$.

\begin{table*}
  \caption{Paths with maximum probability for the terminal configurations
           of neon.
           Here ``Config.''~is the cationic configuration, ``Paths''~is
           the number of paths leading to the configuration, ``Unique''~is
           the number of unique strings, ``Norm.''~is
           the number of unique normalized strings, ``String''~is
           the string of the path with maximum probability~``Probability'',
           and ``Term.~prob.''~is the terminal probability in the configuration.
           The empty string is~$\varepsilon$~\cite{Nagornyi:AL-16}.
           The photon energy is~$1200 \eV$, the fluence
           is~$8 \E{11} \U{\tfrac{photons}{\mu m^2}}$, and the
           pulse duration (FWHM) is~$100 \U{fs}$.
           ``Paths'' counts also paths which have zero weight because
           of the photon energy being too low for the involved
           photoionization channels.}
  \begin{ruledtabular}
    \begin{tabular}{crrrlrr}
      Config. & \hfill Paths\hfill\hfill & Unique & Norm. & String &
      Probability & Term.~prob. \\
      \hline
      226 &      1 &     1 &   1 & $\varepsilon$ & $2.8 \E{-7}$ & $2.8 \E{-7}$ \\
      225 &      2 &     2 &   2 & V             & $9.5 \E{-7}$ & $9.5 \E{-7}$ \\
      224 &      9 &     8 &   6 & CA            & $4.2 \E{-5}$ & $4.2 \E{-5}$ \\
      223 &     45 &    30 &   8 & VCA           & $3.8 \E{-5}$ & $7.7 \E{-5}$ \\
      222 &    237 &   115 &  18 & CACA          & $5.8 \E{-4}$ & $6.2 \E{-4}$ \\
      221 &   1261 &   441 &  22 & VCACA         & $2.3 \E{-4}$ & $8.0 \E{-4}$ \\
      220 &   6719 &  1693 &  41 & CACACA        & $2.8 \E{-3}$ & $3.5 \E{-3}$ \\
      210 &  27403 &  4128 &  41 & VCACACA       & $1.2 \E{-3}$ & $7.9 \E{-3}$ \\
      200 & 185016 & 19067 & 119 & CACACACA      & $2.0 \E{-2}$ & $3.1 \E{-1}$ \\
      100 & 574522 & 58364 & 175 & CACACACAC     & $4.3 \E{-2}$ & $6.8 \E{-1}$ \\
      000 & 806829 & 81803 & 175 & VCACACDVVV    & $2.2 \E{-7}$ & $3.1 \E{-6}$ \\
    \end{tabular}
  \end{ruledtabular}
  \label{tab:maxprob}
\end{table*}

Equipped with the paths and the branching ratios, we assign weights
to the paths.
The path of length one, the neutral atom, has unit weight.
To find the weight of a path with a certain length greater than one, we
multiply the weight of the path of length minus one by
the fraction~$\tfrac{\sigma_{j \leftarrow i}}{\sigma_i} \, B_{\mathrm{P}, i}$,
for a photoionization
and by~$\tfrac{\Gamma_{k, j \leftarrow i}}{\Gamma_{k, i}} \, B_{k, i}$
for a decay with~$k \in \{\mathrm{E}, \mathrm{R}\}$ where
the last configuration of the path with length minus one is~$i$
and the last configuration of the path with length is~$j$.
Sums of the path weights for all configurations of a fixed charge state---along
horizontal lines in the graph~\cite{Kozyrev:GR-16}
in~\fref{fig:req_graph}---give unity provided that no paths are omitted
and double counting is removed:
the first condition refers to our recursive algorithm~\cite{Uspenskii:AL-11}
which omits radiative transitions if a competing Auger decay is present;
the second condition implies that paths with trailing radiative transitions
are disregarded in the sum over the weights.
The fact that the sum of the path weights under the previous conditions
is unity is because starting from the neutral atom with unit probability,
the paths leading to configurations of a chosen charge state reveal how this
unit probability diffuses over these configurations.
But the paths leading to non-terminal configurations have no
relevance for our analysis because no probability remains in
those configurations for~$t \to \infty$---\ie, the so-called terminal
probability vanishes for these configurations---as they are connected by
radiative transitions to terminal configurations.
Hence these paths are suppressed altogether.

To convert the weights of the paths into probabilities, we divide the
weight of a path ending in a terminal configuration by the sum of all
the weights of the paths that lead to this terminal configuration multiplied by
the terminal probability of that configuration.
Paths that lead to non-terminal configurations consequently have zero
probability.
This assigns a probability to each path that indicates its contribution
to the probability found in the configuration at its last vertex
for~$t \to \infty$.

\subsection{Formal language and summed quantities}

With the probabilities of the paths from the previous~\sref{sec:probabilities},
we figure out which paths are most important to reach a chosen
configuration.
For neon, we have a total number of paths of~$2 \, 051 \, 874$
and $1 \, 602 \, 044$~many paths lead to its terminal configurations.
In order to symbolize paths, we introduced the
alphabet~\cite{Nagornyi:AL-16}~``ACV'' at the beginning
of~\sref{sec:probabilities}.
Thereby, we disregard a distinction of ionization from~$2s$ or $2p$~subshells
of neon and summarize such an ionization under the letter~``V''.
We would like to extend the alphabet to include photoionization of the
remaining core electron of a single core hole producing a double core hole~``D''
and radiative decay~``R''.
With this alphabet~``ACDRV'', we can form strings (or
words)~\cite{Nagornyi:WO-16} that represent the transitions in a path.
A path of a certain length has length minus one transitions, \ie,
it is described by a string of length minus one.
The path starting and ending in the neutral atom is represented
by the empty string~$\varepsilon$~\cite{Nagornyi:AL-16}.
This identification allows us to understand the algorithm~\cite{Uspenskii:AL-11}
to generate the paths in the graph~\cite{Kozyrev:GR-16}
of~\fref{fig:req_graph} as a formal grammar~\cite{Gladkii:GF-11}
that specifies the words~\cite{Nagornyi:WO-16},
\ie, the strings, over the given alphabet~\cite{Nagornyi:AL-16} of the
formal language~\cite{Gladkii:FL-16}.
This approach already reduces the intricacy of the problem.
Yet we still have many paths which does not reveal interesting information.

First, we determine the path with maximum probability for each terminal
configuration in~\tref{tab:maxprob}.
As we thus select only a single path per charge state, the
associated values in the column ``Probabilities'' in~\tref{tab:maxprob}
do not sum to unity.
However, because the atom is in some terminal configuration after the
interaction, the values in the column ``Term.~prob.'' do sum to unity.
We find that, indeed, for the specified pulse parameters
``CA''\nbh{}style processes dominate as proposed in
Refs.~\onlinecite{Rohringer:XR-07,Young:FE-10}.
Valence ionization is only present, if the terminal configuration
cannot be reached by ``CA''\nbh{}style processes, \ie, an odd number of
electrons is missing or the \xray~energy is too low for~``C'' to occur.
Comparing the probabilities of the paths with maximum probability
with the terminal probabilities reveals that even for the configuration~$225$
there is no agreement and the former is lower than the latter if
more digits are considered than shown in~\tref{tab:maxprob}.
Namely, as the sum of the probabilities of all paths that lead to a terminal
configuration is the terminal probability, there needs to be an
additional path that accounts for the discrepancy.
In this case, it is ``VR'', \ie, valence ionization of~226 to~216
with ensuing radiative decay to~225.
Yet for higher charged configurations, it turns out that the path
with maximum probability contributes only a small amount to the
terminal probability trapped in a configuration [\tref{tab:maxprob}].

\begin{table}
  \caption{Strings with a probability greater than~$10^{-5}$ for
           selected terminal configurations of neon.
           Here ``Probability''~is the sum of the probabilities of all
           paths with the same ``String'' and ``Sum''~is the sum of the
           probabilities of the strings shown for a configuration in the table.
           Other parameters as in~\tref{tab:maxprob}.}
  \begin{ruledtabular}
    \begin{tabular}{clrr}
       Config. & String & Probability & \hfill Sum\hfill\  \\
      \hline
      223 & VCA    & $3.8 \E{-5}$ & $7.7 \E{-5}$ \\
          & CAV    & $3.4 \E{-5}$ \\
      222 & CACA   & $5.8 \E{-4}$ & $6.2 \E{-4}$ \\
          & CDAA   & $3.2 \E{-5}$ \\
      221 & VCACA  & $2.3 \E{-4}$ & $7.7 \E{-4}$ \\
          & CAVCA  & $2.1 \E{-4}$ \\
          & CACAV  & $1.8 \E{-4}$ \\
          & CACVA  & $5.0 \E{-5}$ \\
          & CVACA  & $3.3 \E{-5}$ \\
          & VCDAA  & $2.1 \E{-5}$ \\
          & VCACAR & $1.9 \E{-5}$ \\
          & CAVCAR & $1.7 \E{-5}$ \\
          & CACAVR & $1.3 \E{-5}$ \\
    \end{tabular}
  \end{ruledtabular}
  \label{tab:largestprob}
\end{table}

Second, for configurations~$224, \ldots, 000$ multiple paths with the
same string arise that contribute comparatively to the terminal probability.
This is due to the reduced level of detail by introducing letters for processes
which summarize different configurations under a single letter.
As there is no distinction made between paths with the same string,
we sum over their probabilities and thus remove these duplicates.
This significantly reduces the number of strings to be considered
[\tref{tab:maxprob}].
Taking into account the strings with highest probability,
\tref{tab:largestprob}, provides an understanding of the terminal
configurations~$223$, $222$, and $221$.
For~$223$ there are two dominant strings, ``VCA'' and ``CAV'', which make
up most of the probability.
The~``CAV'' has a slightly lower probability than ``VCA'' because there
are fewer valence electrons available for the ``V''~transition.
The string~``CVA'' has a much smaller probability as, in this case, the
short core-hole lifetime sets a time scale during which a valence
electron needs to be ionized.
For~$222$ and $221$~we find a small admixture of strings with double core
holes.
Finally, for $221$~several strings contribute comparatively which are
distinguished by the position of the ``V''.
For higher charge states, too many strings make a noticeable
contribution to the terminal probability for an analysis using the strings
so far introduced to be conclusive~\cite{SuppData}.

\begin{table}
  \caption{Normalized strings with a probability in the highest two
           orders of magnitude for the probability of selected terminal
           configurations of neon.
           Here ``Norm. string''~is a normalized string and ``Probability''~is
           the sum over the probabilities of all paths with the same
           normalized string.
           The ``Sum'' is the sum of the probabilities of the normalized
           strings shown for a configuration in the table.
           Other parameters as in~\tref{tab:maxprob}.}
  \begin{ruledtabular}
    \begin{tabular}{clrr}
       Config. & Norm. string & Probability & \hfill Sum\hfill\  \\
      \hline
      221 & AACCV      & $7.0 \E{-4}$ & $8.0 \E{-4}$ \\
          & AACCRV     & $5.7 \E{-5}$ \\
          & AACDV      & $3.8 \E{-5}$ \\
      220 & AAACCC     & $2.8 \E{-3}$ & $3.4 \E{-3}$ \\
          & AAACCD     & $4.0 \E{-4}$ \\
          & AAACCCR    & $2.6 \E{-4}$ \\
      210 & AAACCCV    & $6.0 \E{-3}$ & $7.7 \E{-3}$ \\
          & AAACCDV    & $1.1 \E{-3}$ \\
          & AAACCCRV   & $6.1 \E{-4}$ \\
      200 & AAAACCCC   & $1.6 \E{-1}$ &2.9 $ \E{-1}$ \\
          & AAAACCCD   & $5.5 \E{-2}$ \\
          & AAACCCCRV  & $5.1 \E{-2}$ \\
          & AAACCCDRV  & $1.9 \E{-2}$ \\
      100 & AAAACCCCC  & $3.5 \E{-1}$ & $6.6 \E{-1}$ \\
          & AAAACCCCD  & $1.2 \E{-1}$ \\
          & AAACCCCCRV & $1.1 \E{-1}$ \\
          & AAACCCCDRV & $4.1 \E{-2}$ \\
          & AAACCCCVV  & $2.5 \E{-2}$ \\
          & AAAACCCDD  & $1.1 \E{-2}$ \\
      000 & AACCCDVVVV & $2.5 \E{-6}$ & $3.0 \E{-6}$ \\
          & AACCDDVVVV & $4.7 \E{-7}$ \\
    \end{tabular}
  \end{ruledtabular}
  \label{tab:normstr}
\end{table}

Third, for the highest-charged configurations~$210, \ldots, 000$,
there are so many paths with different strings to reach them that
it is hard to extract a few strings which are dominant.
Hence, we take another step to simplify the problem and introduce
normalized strings.
These are the previously introduced strings whose characters we sort
in the order they appear in the latin alphabet.
Then we may sum the probabilities for the same normalized strings which
greatly reduces the number of strings to consider [\tref{tab:maxprob}].
This facilitates to extract a few important ionization pathways listed
in~\tref{tab:normstr}.
Even for the highest charge states, the number of normalized strings
of relevance remains manageable.
We note an increasing relevance of paths with double core holes for
higher charge states.

\begin{table}
  \caption{Length of strings with a summed probability in the highest three
           orders of magnitude of the probability for the terminal
           configurations of neon.
           Here ``Min.''~refers to the minimum string length possible,
           ``Max.''~to the maximum string length,  and ``Len.''~to the
           string length associated with ``Probability''.
           The ``Sum'' is the sum of the probabilities of the
           string lengths shown for a configuration in the table.
           Other parameters as in~\tref{tab:maxprob}.}
  \begin{ruledtabular}
    \begin{tabular}{crrrcc}
       Config. & Min. & Max. & Len. & Probability & Sum \\
      \hline
      226 &  0 &  0 &  0 & $2.8 \E{-7}$ & $2.8 \E{-7}$ \\
      225 &  1 &  2 &  1 & $9.5 \E{-7}$ & $9.6 \E{-7}$ \\
          &    &    &  2 & $1.2 \E{-9}$ \\
      224 &  2 &  4 &  2 & $4.2 \E{-5}$ & $4.2 \E{-5}$ \\
      223 &  3 &  6 &  3 & $7.7 \E{-5}$ & $7.7 \E{-5}$ \\
          &    &    &  4 & $3.4 \E{-7}$ \\
      222 &  4 &  8 &  4 & $6.2 \E{-4}$ & $6.2 \E{-4}$ \\
          &    &    &  5 & $1.6 \E{-6}$ \\
      221 &  5 & 10 &  5 & $7.4 \E{-4}$ & $8.0 \E{-4}$ \\
          &    &    &  6 & $6.0 \E{-5}$ \\
      220 &  6 & 12 &  6 & $3.2 \E{-3}$ & $3.5 \E{-3}$ \\
          &    &    &  7 & $3.1 \E{-4}$ \\
      210 &  7 & 13 &  7 & $7.2 \E{-3}$ & $7.9 \E{-3}$ \\
          &    &    &  8 & $7.2 \E{-4}$ \\
      200 &  8 & 15 &  8 & $2.3 \E{-1}$ & $3.1 \E{-1}$ \\
          &    &    &  9 & $7.6 \E{-2}$ \\
          &    &    & 10 & $1.4 \E{-3}$ \\
      100 &  9 & 16 &  9 & $5.1 \E{-1}$ & $6.8 \E{-1}$ \\
          &    &    & 10 & $1.6 \E{-1}$ \\
          &    &    & 11 & $3.0 \E{-3}$ \\
      000 & 10 & 17 & 10 & $3.0 \E{-6}$ & $3.1 \E{-6}$ \\
          &    &    & 11 & $3.2 \E{-8}$ \\
    \end{tabular}
  \end{ruledtabular}
  \label{tab:strlen}
\end{table}

Forth, the most drastic simplification of the problem is to disregard
the types of transitions occurring and to consider
only the length of the strings of the paths leading to a terminal configuration
and sum all probabilities of strings with this length.
From~\tref{tab:strlen}, we realize that this simplification leads
to only one, two, or three string lengths that make a substantial contribution
to the probability of a terminal configuration.
The minimum string length to reach a specific charge state is, thereby,
given by the number of electrons that need to be removed.
It turns out, \tref{tab:strlen}, that this already leads to the paths
which make the most important contribution.
Electrons are removed by ``A'', ``C'', ``D'', ``V''~transitions whereas
``R''~transitions leave the electron number unchanged.
Thus strings longer than the minimum length contain ``R''~transitions
which are slow compared with ``A''~transitions and, at high \xray~intensity,
with ``C'', ``D'', ``V''~transitions [\fref{fig:sch_Auger_widths}].
The maximum string length specified in~\tref{tab:strlen} is somewhat
shorter than it could be because we neglected radiative transitions
if a competing Auger decay is present.

\section{Computational details}
\label{sec:compdet}
\subsection{Complete atomic electronic structure for rate equations}
\label{sec:CAESR}

To carry out the atomic electronic structure computations from~\sref{sec:AES},
we developed the program \textsc{caesr} (Complete Atomic Electronic Structure
for Rate equations) as part of the \textsc{fella}
package~\cite{fella:pgm-V1.4.0}.
This program generates all configurations and invokes
\textsc{Grasp2K}~\cite{Jonsson:2K-07,Jonsson:2K-13} and
\textsc{Ratip}~\cite{Fritzsche:RA-12} with the appropriate input files.
Afterwards \textsc{caesr} analyses the output and processes the results.

We use the version~1 programs of the year~2013~edition
of~\textsc{Grasp2K}~\cite{Jonsson:2K-07,Jonsson:2K-13}.
In the computations, we use an atomic mass number of~$20$ for neon
and a mass of the nucleus of~$20.1797 \U{u}$~\cite{Wieser:AW-11}.
Thereby, we assume the most abundant isotope of neon which has a
vanishing nuclear spin, nuclear dipole moment, and
nuclear quadrupole moment~\cite{Stone:TN-14}.
For less than 50~energy levels (always the case for neon), \textsc{caesr}
performs an extended-optimal-level~(EOL) computation~\cite{Froese:MC-16} which,
however, becomes unreliable due to singular energy levels beyond this threshold.
Above 50~levels, \textsc{caesr} switches to an extended-average-level~(EAL)
computation~\cite{Froese:MC-16}.
The EOL~scheme is more suited for some cases with few electrons for which
the EAL~scheme does not converge.
We require a minimum amount of orbital convergence of~$10^{-5}$ in
MCDHF~computations.

For the \textsc{Ratip}~\cite{Fritzsche:RA-12} programs, \textsc{caesr}
initially determines the states of the multiplet for each cationic
configuration of an atom apart from the electron-bare nucleus.
This is accomplished by representing the Hamiltonian in the basis
of the configuration state functions of the multiplet and diagonalizing
it fully with \textsc{Relci}~\cite{Fritzsche:CI-02,Fritzsche:RA-12}.
The Hamiltonian, thereby, comprises the contributions of the
frequency-independent Breit interaction, the vacuum polarization,
and the specific mass shift.
\textsc{Reos} is used to determine radiative transitions in the Babushkin gauge;
it relies on a complete expansion of the atomic state functions in a
determinant basis with~\textsc{Cesd}.
No orthogonality is assumed in \textsc{Reos} between the orbital sets
of the initial and final atomic states.
For neon the lowest multipole order of the radiative transition
is~E$_1$ on configuration level, if an electron falls down
from a $p$~subshell to an $s$~subshell and M$_1$ and E$_2$~for
transitions involving the multiplets of configurations
differing by two $s$~subshells.
For atoms with a higher atomic number than neon, more cases arise.
If~M$_1$ and E$_2$ are the lowest order multipoles, we determine
both radiative decay widths and add them.
\textsc{Auger} and \textsc{Photo} are employed to calculate the
electronic decay widths and photoionization cross sections, respectively.
In both cases orthogonality between the sets of orbitals of the initial and
final atomic states is assumed.
Also a continuum electron is emitted for both quantities;
its wave function is determined including exchange interactions with
the parent ion for \textsc{Auger} and without it for \textsc{Photo}.
The photoionization cross sections are computed in Babushkin gauge
in E$_1$~multipole order on a linear-logarithmic grid of energies starting
at the ionization threshold of the atom;
the first 10~energies are linearly spaced with a step size of~$1 \eV$;
the remaining 20~energies are spaced evenly on a decadic-logarithmic grid
starting~$1 \eV$ above the last energy of the linear grid and ending
at~$2000 \eV$.
This choice is motivated by the observation that cross sections vary most
close to the ionization threshold and very little afterwards.

\subsection{Numerical solution of the rate equations}

The \xray~pulse is represented temporally by a Gaussian function
[Eq.~(7) in Ref.~\onlinecite{Buth:UA-12}];
there is no need to account for the spikiness of SASE pulses from
LCLS~\cite{Rohringer:XR-07,Buth:UA-12}.
Spatially, \xray~pulses are described as a Gaussian beam
[Eq.~(8) in Ref.~\onlinecite{Buth:UA-12}] with a
cross section that has a specific FWHM major and minor axes;
longitudinally, the beam has a long Rayleigh length and thus a small
variation over the interaction volume, \ie, a longitudinal variation
is neglected~\cite{Buth:UA-12}.

We numerically solve the rate equations [\sref{sec:xrayint}]
in the time interval~$[-10 \U{ps}; 10 \U{ps}]$ which is long enough for the
\xray~pulse to be over and all Auger decay processes to have taken place.
However, we would have to solve the rate equations for an excessively long
time interval for radiative transitions to take place where no competing
Auger decay is available because the lifetime of radiative decay
is so much longer than the one of Auger decay [\fref{fig:sch_Auger_widths}].
This is particularly the case for electric-dipole-forbidden transitions.
The slowest radiative decay is~$110 \to 200$ with a lifetime of~$134 \U{\mu s}$
and the slowest Auger decay is~$111 \to 200$ with a lifetime of~$41.7 \U{fs}$.
The relevant radiative transitions are the ones originating
from the configurations [Eq.~\eref{eq:configurations}]:
\begin{eqnarray}
  \label{eq:radconfig}
  \mathbb D = \{&&216, 215, 206, 214, 205, 213, 204, 212, \\
    &&203, 211, 202, 201, 110, 101, 010, 001\} \subset \mathbb K \; . \nonumber
\end{eqnarray}

After the \xray~pulse is over, only decay processes continue to occur.
Hence we proceed analogously to Theorem~2 in Ref.~\onlinecite{Buth:LT-18},
\ie, we decompose the solution of the rate equations
into a numerical solution up to time~$\mathfrak T$ and a
subsequent evolution in terms of the decay equation~\cite{Buth:LT-18}, \ie,
Eq.~\eref{eq:reqgam} without terms involving cross sections,
\begin{equation}
  \label{eq:decayeqn}
  \dfrac{\differential \vec{\mathfrak p}(t)}{\differential t} = \mat \Gamma
    \; \vec {\mathfrak p}(t) \; ,
\end{equation}
yielding~$\vec{\mathfrak p}(t) = \euler^{\mats \Gamma \, (t - \mathfrak T)} \>
\vec{\mathfrak p}(\mathfrak T)$.
Here the $|\mathbb K|$-many probabilities are aggregated
in~$\vec {\mathfrak p}(t)$ where the initial
value~$\vec {\mathfrak p}(\mathfrak T)$ is
given by the probabilities from the numerical solution at time~$\mathfrak T$.
In our case, $\mat \Gamma$~comprises only the radiative decay widths
originating from the configurations~\eref{eq:radconfig}.

If we were only interested in the terminal probabilities,
then we would not need to solve~\eref{eq:decayeqn}.
Instead, we only needed to sum over the probabilities of each charge state
at time~$\mathfrak T$;
the sums are then the terminal probabilities.
This is what comes out, if the limit~$t \to \infty$ is taken
in~\eref{eq:eigprob} below ($\mu_{\ell} \leq 0$).
However, numerically integrating~\eref{eq:decayprob} ($k = \mathrm{R}$)
and \eref{eq:photonem} does require the solution of the
rate equations for an excessively long time interval.
To accomplish this, we notice that~$\mat \Gamma$ is diagonalizable,
and go to its eigenbasis~\cite{Buth:LT-18}
which is a representation of the solution of Eq.~\eref{eq:decayeqn}
for the probability~$\mathfrak p_i(t)$ with~$i \in \mathbb K$
that is more suited for our purposes:
\begin{equation}
  \label{eq:eigprob}
  \mathfrak p_i(t) = \Sum_{j,\ell \in \mathbb K} \mathfrak U_{i\ell} \;
    \euler^{\mu_{\ell} \, (t - \mathfrak T)} \,
    (\mat{\mathfrak U}^{-1})_{\ell j} \> \mathfrak p_j(\mathfrak T) \; .
\end{equation}
Here $\mu_{\ell}$ and $\mat{\mathfrak U}$ denote the
eigenvalue for~$\ell \in \mathbb K$ and the matrix of eigenvectors
of~$\mat \Gamma$, respectively.
To numerically integrate~\eref{eq:decayprob} ($k = \mathrm{R}$)
and \eref{eq:photonem}, we need to compute~$\Int_{\tau}^{\mathfrak T} P_i(t')
\differential t' + \Int_{\mathfrak T}^t \mathfrak p_i(t') \differential t'$
where $\tau = -10 \U{ps}$ here and $t \geq \mathfrak T = 10 \U{ps}$.
The second integral is solved with~\eref{eq:eigprob} using
\begin{equation}
  \label{eq:expint}
  \Int_{\mathfrak T}^t \euler^{\mu_{\ell} \, (t' - \mathfrak T)}
    \differential t' =
  \begin{cases}
    \tfrac{\euler^{\mu_{\ell} \, (t - \mathfrak T)} - 1}{\mu_{\ell}} &
      \mu_{\ell} \neq 0 \\
    t - \mathfrak T & \mu_{\ell} = 0 \; .
  \end{cases}
\end{equation}
By replacing~$\euler^{\mu_{\ell} \, (t - \mathfrak T)}$ in~\eref{eq:eigprob}
with~\eref{eq:expint}, we obtain~$\Int_{\mathfrak T}^t \mathfrak p_i(t')
\differential t'$.
We arrive at the probability for photon emission
\begin{equation}
  \label{eq:fluorprob}
  \Lim_{t \to \infty} F_{j \leftarrow i}(t) = F'_{j \leftarrow i}(\mathfrak T)
    + \Lim_{t \to \infty} \Bigl( \mathbf{1}_{\mathbb D}(i) \,
    \Gamma_{\mathrm{R}, j \leftarrow i} \Int_{\mathfrak T}^t
    \mathfrak p_i(t') \differential t' \Bigr) \; .
\end{equation}
Here, $F'_{j \leftarrow i}(\mathfrak T)$~stands for the numerical integration
of~\eref{eq:photonem} in the interval~$[\tau ; \mathfrak T]$ and
$\mathbf{1}_{\mathbb D}(i)$~is the indicator function~\cite{Konyushkov:CF-11}
of~$\mathbb D \subset \mathbb K$ which is unity for~$i \in \mathbb D$
and zero otherwise.
If $i \in \mathbb D$ and $\Gamma_{\mathrm{R}, j \leftarrow i} \neq 0$ then
the integral~$\Int_{\mathfrak T}^{\infty} \mathfrak p_i(t') \differential t'$
is finite because $\mathfrak p_i(t')$~decreases exponentially toward zero.
Otherwise the term in parentheses in~\eref{eq:fluorprob} vanishes.
The integral~\eref{eq:decayprob} ($k = \mathrm{R}$) is solved analogously.

\end{document}